\documentclass[amsmath,amssymb,pre,preprint]{revtex4}

\usepackage{graphicx}
\usepackage{dcolumn}
\usepackage{bm}

\begin{document}

\title{Influence of an external magnetic field on forced turbulence in a swirling flow of liquid metal}
\author{Basile Gallet}
\author{Michael Berhanu}
\author{Nicolas Mordant}

\affiliation{Laboratoire de Physique Statistique, Ecole Normale Sup\'erieure \& CNRS, 24 Rue Lhomond, 75231 PARIS Cedex 05, France}

\date{\today}
\begin{abstract}
We report an experimental investigation on the influence of an external magnetic field on forced 3D turbulence of liquid gallium in a closed vessel. We observe an exponential damping of the turbulent velocity fluctuations as a function of the interaction parameter $N$ (ratio of Lorentz force over inertial terms of the Navier-Stokes equation). The flow structures develop some anisotropy but do not become bidimensional. From a dynamical viewpoint, the damping first occurs homogeneously over the whole spectrum of frequencies. For larger values of $N$, a very strong additional damping occurs at the highest frequencies. 
However, the injected mechanical power remains independent of the applied magnetic field. The simultaneous measurement of induced magnetic field and electrical potential differences shows a very weak correlation between magnetic field and velocity fluctuations. The observed reduction of the fluctuations is in agreement with a previously proposed mechanism for the saturation of turbulent dynamos and with the order of magnitude of the Von K\'arm\'an Sodium dynamo magnetic field.

\end{abstract}


\maketitle

Situations were a magnetic field interacts with a turbulent flow are found in various domains of physics, including molten metals processing, laboratory flows, and astrophysics. The motion of electrically conducting fluid in a magnetic field induces electrical currents, which in turn react on the flow through the Lorentz force. The power injected in the flow is thus shared between two dissipative mechanisms: viscous friction and ohmic dissipation of the induced currents. On the one hand, the situation where the flow is laminar is very well understood, and the geometry of the velocity field and of the induced currents can be computed analytically. On the other hand, several questions remain open in the fully turbulent situation: when a statistically steady state is reached, is the mean injected power higher or lower than in the nonmagnetic case? What controls the ratio of ohmic to viscous dissipation? How is the turbulent cascade affected by the magnetic field? To adress some of these questions, we have designed an experimental device which allows to apply a strong magnetic field on a fully turbulent flow.
\newline

When an electrically conducting fluid is set into motion, a magnetic Reynolds number $Rm$ can be defined as the ratio of the ohmic diffusive time to the eddy-turnover time. This number reaches huge values in galactic flows, but can hardly exceed one in a laboratory experiment. The common liquids of high electrical conductivity (gallium, mercury, sodium) have a very low kinematic viscosity: their magnetic Prandtl number $Pm$ (ratio of the kinematic viscosity over the magnetic diffusivity) is less than $10^{-5}$. A flow with $Rm$ of order one is turbulent and thus requires a high power input to be driven. For this reason most experimental studies have been restricted to low $Rm$. They were conducted mostly in channel flows and grid generated turbulence \cite{Eckert,Alemany}.
The most general observation is that the application of a strong magnetic field leads to a steeper decay of the power spectra of the turbulent velocity fluctuations at large wavenumbers: the decay goes from a classical $k^{-5/3}$ scaling without magnetic field to $k^{-3}$ or $k^{-4}$ for the highest applied fields. In the meantime, some anisotropy is developed leading to larger characteristic scales along the applied magnetic field. The phenomenology of these transformations is quite well understood in terms of the anisotropy of the ohmic dissipation \cite{Knaepen}. One difficulty arises from the fact that the boundary conditions can have a strong effect on the turbulence level. In channel flows the choice between conducting and insulating walls strongly impacts the flow. A strong external magnetic field perpendicular to the boundaries leads to an increase of the turbulence level due to modifications of the boundary layers \cite{Eckert}. On the contrary, Alemany et al. observed in grid generated decaying turbulence an enhanced decay of the turbulent fluctuations. As far as forced turbulence is concerned, Sisan et al. studied the influence of a magnetic field on a flow of liquid sodium inside a sphere \cite{Sisan}. However, the flow again goes through a variety of instabilities which prevents a study solely focussed on the impact of the magnetic field on the turbulent fluctuations: the geometry of the mean flow and of the induced magnetic field keeps changing as the magnetic field is increased.

 Several numerical simulations of this issue have been conducted \cite{Vorobev, Zikanov, Zikanov2,Burattini}. Despite the rather low spatial resolution available for this problem, these works show the same phenomenology: development of anisotropy, steepening of the spectra and trend to bidimensionalization of the flow as the magnetic field increases. The direct numerical simulations (DNS) also allow to compute the angular flux of energy from the energy-containing Fourier modes (more or less orthogonal to the applied field) to the modes that are preferentially damped by ohmic dissipation. Once again, the magnetic  field can affect the large scale structure of the velocity field: the interaction between forcing and magnetic field in the DNS of Zikanov \& Thess~\cite{Zikanov, Zikanov2} leads to an intermittent behavior between phases of roughly isotropic flow and phases of bidimensional columnar vortices that eventually get unstable. Vorobev et al.\cite{Vorobev} and Burratini et al.\cite{Burattini} performed DNS in a forced regime which are closely related to our experiment. However, the maximum kinetic Reynolds number reached in their studies remains orders of magnitude below what can be achieved in the laboratory. An experimental investigation at very high kinetic Reynolds number thus remains necessary to characterize the effect of a strong magnetic field on a fully developped turbulent cascade.

The flow under study in our experiment resembles the Von K\'arm\'an geometry: counter-rotating propellers force a flow inside a cylindrical tank. The large scale flow is known to have a strong shear layer in the equatorial plane where one observes the maximum level of turbulence \cite{Marie}. We do not observe any bifurcation of the flow as the magnetic field increases nor any drastic change of the large scale recirculation imposed by the constant forcing. These ingredients strongly differ from the previously reported experiments and allow to study the influence of an applied magnetic field on the turbulent cascade in a given flow geometry. This issue arises in the framework of turbulent laboratory dynamo studies such as the Von K\'arm\'an Sodium (VKS) experiment \cite{Monchaux} which flow's geometry is similar to ours. In the turbulent dynamo problem, one issue is to understand the precise mechanism for the saturation of the magnetic field. Once the dynamo magnetic field gets strong enough, it reacts on the flow through the Lorentz force. This usually reduces the ability of the flow to sustain dynamo action. An equilibrium can be reached, so that the magnetic field saturates. This backreaction changes the properties of the bulk turbulence. In a situation where an $\alpha$ effect takes part in the generation of the dynamo, i.e. if the turbulent fluctuations have a mean field effect, then the changes in the statistics of turbulence should be involved in the saturation mechanism of the dynamo.

\section{Description of the experiment}

\subsection{The flow}

\begin{figure}[]
\includegraphics[width=8 cm]{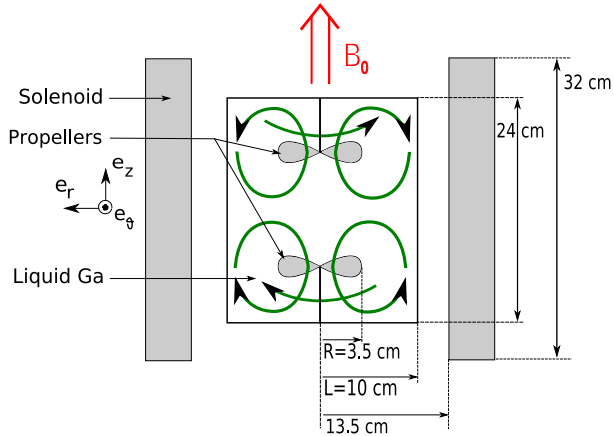}
\caption{\label{setup}Sketch of the experimental setup. The curved arrows represent the average large-scale motion of the fluid.}
\end{figure}
Our turbulent flow resembles the Von K\'arm\'an geometry, which is widely used in experiments on turbulence and magnetohydrodynamics (\cite{Monchaux, Marie} for example). 8 liters of liquid gallium are contained in a closed vertical cylinder of diameter 20~cm and height 24~cm. The thickness of the cylindrical wall is 7.5~mm, and that of the top and bottom walls is 12~mm (Fig.~\ref{setup}). This tank is  made of stainless steel. Gallium melts at about 30$^\circ C$. Its density is $\rho=6090$~kg/m$^3$ and its kinematic viscosity is $\nu=3.11\, 10^{-7}$~m$^2$/s. Its electrical conductivity at our operating temperature ($45^\circ$C) is $\sigma=3.9\,10^6$~$\Omega^{-1}$m$^{-1}$. The main difference with the usual Von K\'arm\'an setup is the use of two propellers to drive the flow, rather than impellers or disks. The two propellers are coaxial with the cylinder. They are made of 4 blades inclined 45 degrees to the axis. The two propellers are counter-rotating and the blades are such that both propellers are pumping the fluid from the center of the tank towards its end faces. The large scale flow is similar to the traditional counter-rotating geometry of the Von K\'arm\'an setup.  The propellers are 7 cm in diameter and their rotation rate $f_{rot}$ ranges from $3$~Hz to $30$~Hz. They are entrained at a constant frequency by DC motors, which drives a large scale flow consisting in two parts : first the fluid is pumped from the center of the vessel on the axis of the cylinder by the propellers. It loops back on the periphery of the vessel and comes back to the center in the vicinity of the equatorial plane. In addition to this poloidal recirculation, the fluid is entrained in rotation by the propellers. Differential rotation of the fluid is generated by the counter rotation of the propellers, which induces a strong shear layer in the equatorial plane where the various probes are positioned. The shear layer generates a very high level of turbulence \cite{Marie}: the velocity fluctuations are typically of the same order of magnitude as the large scale circulation. 

A stationary magnetic field $\mathbf B_0$ is imposed by a solenoid. The solenoid is coaxial with the propellers and the cylinder. $\mathbf B_0$ is mainly along the axis of the cylinder with small perpendicular components due to the finite  size of the solenoid (its length is 32~cm and its inner diameter is 27~cm). The maximum magnetic field imposed at the center of the tank is 1600~G. 

\subsection{Governing equations and dimensionless parameters}

The fluid is incompressible and its dynamics is governed by the Navier-Stokes equations:
\begin{eqnarray}
\label{eqNS} \rho\left(\frac{\partial \mathbf v}{\partial t}+(\mathbf v \cdot \nabla)\mathbf v\right) & = & -\nabla p+\rho\nu \Delta \mathbf v + \mathbf j \times \mathbf B \\
\nabla \cdot \mathbf v & = & 0
\end{eqnarray}
where $\mathbf v$ is the flow velocity, $p$ is the pressure, $\mathbf B$ is the magnetic field and $\mathbf j$ is the electrical current density. The last term in the r.h.s. of (\ref{eqNS}) is the Lorentz force.

A velocity scale can be defined using the velocity of the tip of the blades. The radius of the propeller being $R=3.5$~cm, the velocity scale is $2\pi Rf_{rot}$. A typical length scale is the radius of the cylinder $L=10$~cm.  The kinetic Reynolds number is then 
\begin{equation}
Re=\frac{2\pi RL f_{rot}}{\nu}\,.
\end{equation}
At the maximum speed reported here it reaches $2\, 10^6$. The flow is then highly turbulent and remains so even for a rotation rate ten times smaller.

In the approximations of magnetohydrodynamics, the temporal evolution of the magnetic field follows the induction equation:
\begin{equation}
\frac{\partial \mathbf B}{\partial t}=\nabla \times (\mathbf v\times \mathbf B)+\eta\Delta \mathbf B\,.
\end{equation}
Here $\eta=(\mu_0\sigma)^{-1}$ is the magnetic diffusivity,  $\mu_0$ being the magnetic permeability of vacuum ($\eta=0.20$~m$^2$s$^{-1}$ for gallium so that $Pm=\nu/\eta=1.6\,10^{-6}$)  \cite{Moffatt}. The first term in the r.h.s describes the advection and the induction processes. The second term is diffusive and accounts for ohmic dissipation.
A magnetic Reynolds number can be defined as the ratio of the former over the latter:
\begin{equation}
Rm=\mu_0\sigma 2\pi R L f_{rot}\,,
\end{equation}
It is a measure of the strength of the induction processes compared to the ohmic dissipation \cite{Moffatt}. For $Rm\rightarrow 0$ the magnetic field obeys a diffusion equation. For $Rm\gg1$, it is transported and stretched by the flow. This can lead to dynamo action, i.e. spontaneous generation of a magnetic field sustained by a transfer of kinetic energy from the flow to magnetic energy. In our study, $Rm$ is about 3 at the maximum speed reported here ($f_{rot}=30$~Hz): Induction processes are present but do not dominate over diffusion. As $f_{rot}$ goes from 3 to 30~Hz, the magnetic Reynolds number is of order 1 in all cases.

In our range of magnetic Reynolds number, the induced magnetic field $\mathbf b=\mathbf B-\mathbf B_0$ is weak relative to $\mathbf B_0$ (of the order of $1\%$). The Lorentz force is $\mathbf j \times \mathbf B_0$ where $\mathbf j$ is the current induced by the motion of the liquid metal (no exterior current is applied). From Ohm's law, this induced current is $\mathbf j \sim \sigma \mathbf v \times \mathbf B_0$ and is thus of order $2\pi \sigma Rf_{rot}B_0$. One can define an interaction parameter $N$ that estimates the strength of the Lorentz force relative to the advection term in the Navier-Stokes equation:
\begin{equation}
N=\frac{\sigma L B_0^2}{2\pi \rho R f_{rot}}\,,
\label{N}
\end{equation} 
Because of the $1/f_{rot}$ factor, this quantity could reach very high values for low speeds, but the flow would not be turbulent anymore. In order to remain in a turbulent regime, the smallest rotation rate reported here is 3~Hz. The maximum interaction parameter (built with the smallest rotation frequency and the strongest applied magnetic field) is then approximately $2.5$.
Many studies on magnetohydrodynamics use the Hartmann number to quantify the amplitude and influence of the magnetic field. This number measures the ratio of the Lorentz force over the viscous one. It is linked to the interaction parameter by the relation $Ha = \sqrt{N Re}$ which gives the typical value $Ha \simeq 700$ for the present experiment. The influence of the electrical boundary conditions at the end faces is determined by this number and the conductivity ratio $K=\frac{\sigma_w l_w}{\sigma L}$, where $\sigma_w$ is the electrical conductivity of the walls and $l_w$ their thickness. A very detailed numerical study on this issue has been performed in the laminar situation on a flow which geometry is very similar to the present one in \cite{Kharicha}. As far as turbulent flows are concerned,  Eckert et al. stress the importance of the product $K Ha$, which measures the fraction of the electrical current that leaks from the Hartmann layers into the electrically conducting walls. With 12~mm thick end faces $K$ is about $0.05$ so that $K Ha \simeq 35$: the walls have to be considered as electrically conducting. These boundary effects are of crucial importance in experiments where the boundary layers control the rate of turbulence of the flow. However, in the present experimental setup the fluid is forced inertially and we have checked that the magnetic field has only little influence on the mean flow (see section \ref{indmag} below). Furthermore, the high value of the Hartmann number implies that the dominant balance in the bulk of the flow is not between the Lorentz and viscous forces but between the Lorentz force and the inertial term of the Navier-stokes equation: in the following, it is the interaction parameter and not the Hartmann number that leads to a good collapse of the data onto a single curve. For these reasons the present experimental device allows to study the influence of a strong magnetic field on the forced turbulence generated in the central shear layer, avoiding any boundary effect.

\subsection{Probes and measurements}

The propellers are driven at constant frequency. The current provided to the DC motors is directly proportional to the torques they are applying. It is recorded to access the mechanical power injected in the fluid.
\newline

\begin{figure}[!htb]
\includegraphics[width=5 cm]{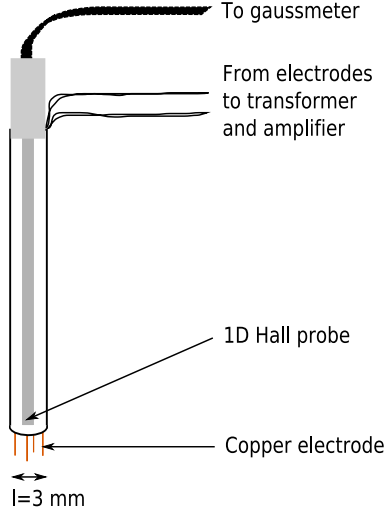}
\caption{\label{pot}Schematics of the potential probes. These probes are vertical and the electrodes are positioned in the mid-plane of the tank, where the shear layer induces strong turbulence.}
\end{figure}
As liquid metals are opaque, the usual velocimetry techniques such as Laser Doppler Velocimetry or Particle Image velocimetry cannot be used. Hot wire anemometry is difficult to implement in liquid metals even if it has been used in the past \cite{Alemany}. Other velocimetry techniques have been developed specifically for liquid metals. Among those are the potential probes. They rely on the measurement of electric potential differences induced by the motion of the conducting fluid in a magnetic field \cite{Ricou, Eckert, Tsinober}. The latter can be applied locally with a small magnet or at larger scale as in our case. We built such probes with 4 electrodes (Fig.~\ref{pot}). The electrodes are made of copper wire, 1~mm in diameter, and insulated by a varnish layer except at their very tip. The electrodes are distant of $l\sim3$~mm. The signal is first amplified by a factor 1000 with a transformer model 1900 from Princeton Applied Research.  It is further amplified by a Stanford Research low noise preamplifier model SR560 and then recorded by a National Instrument DAQ. Note that because of the transformer the average potential cannot be accessed, so that our study focuses on the turbulent fluctuations of the velocity field. 

\begin{figure}[!htb]
(a)\includegraphics[width=8cm]{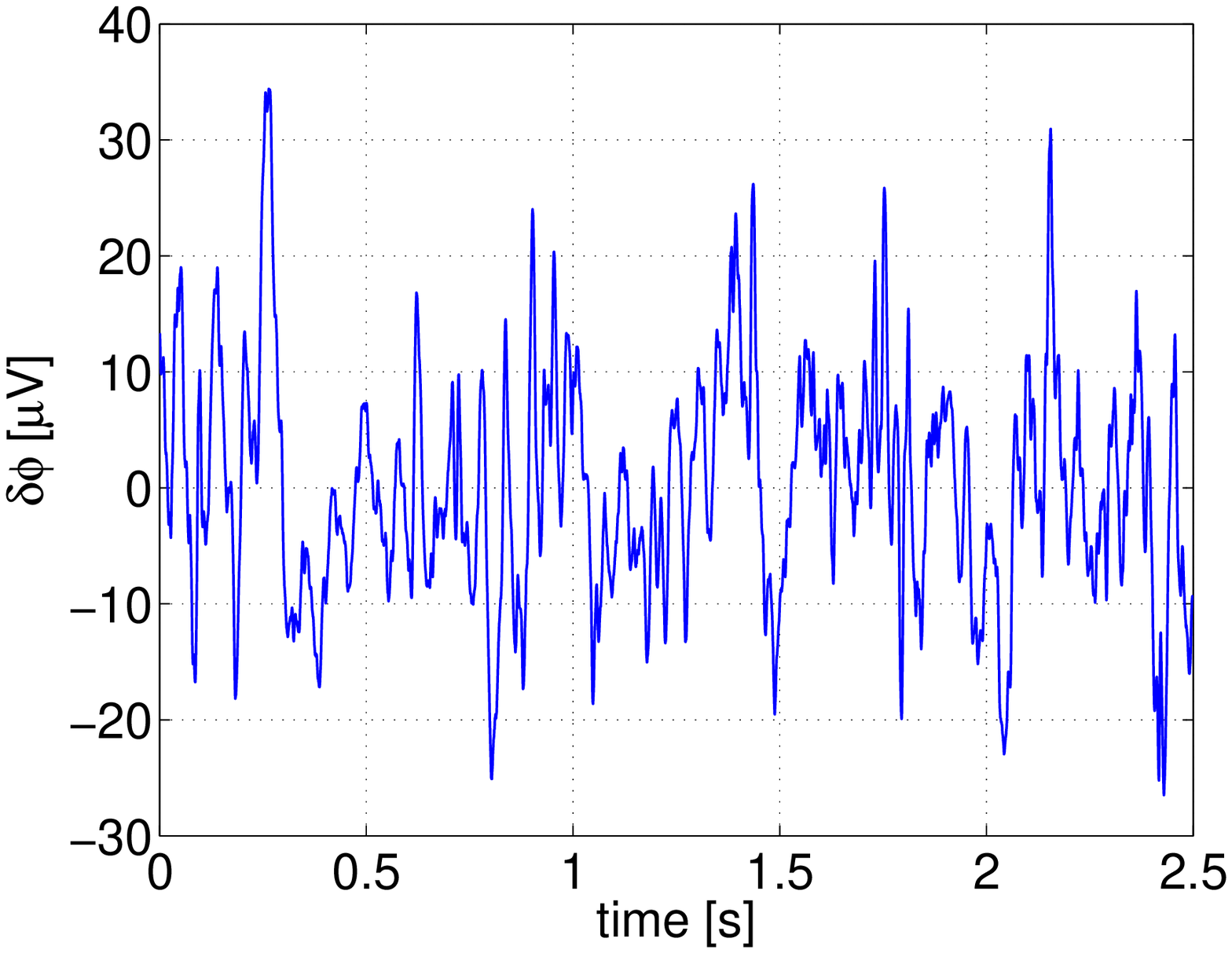}

(b)\includegraphics[width=8cm]{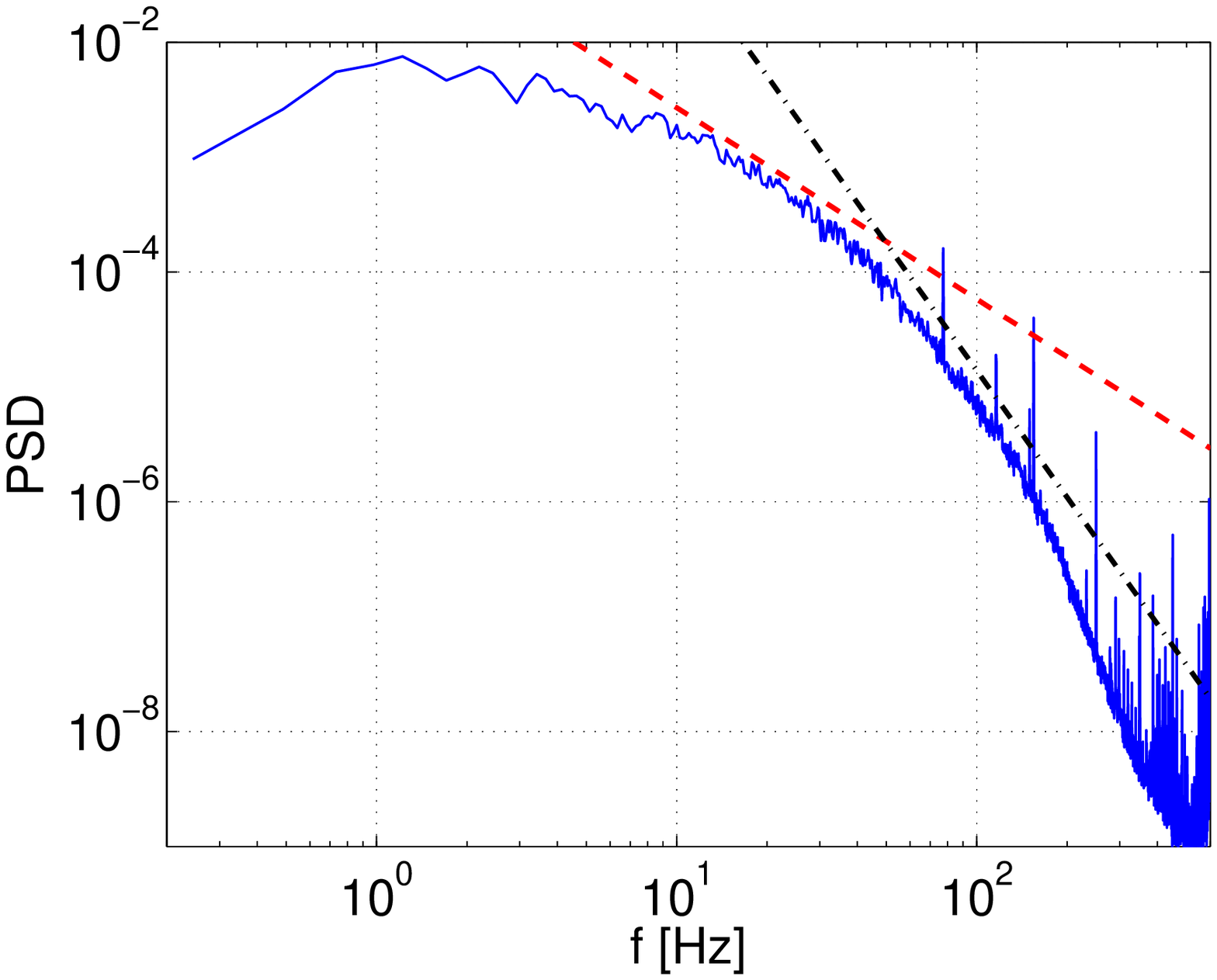}
\caption{\label{sigtemp}(a) Time series from a potential probe for $f_{rot}=20$~Hz and an applied magnetic field of 178~G. (b) Corresponding power spectrum density. The dashed line is a $f^{-5/3}$ scaling and the mixed line a $f^{-11/3}$ scaling. These straight lines are drawn as eye guides.}
\end{figure}
For a steady flow, and assuming $\mathbf j= \mathbf 0$, Ohm's law gives $\nabla \phi=\mathbf v\times \mathbf B$ so that the electric potential difference between the electrodes is directly related to the local velocity of the fluid. One gets $\delta \phi=\int \mathbf v\times \mathbf B \cdot \mathbf dl$ integrating between the two electrodes. If $\mathbf B$ is uniform then $\delta \phi=v_{\bot}Bl$ where $v_{\bot}$ is the component of the velocity orthogonal to both the magnetic field and the electrodes separation. In the general time-dependent case, the link is not so direct. Using Coulomb's gauge and taking the divergence of Ohm's law, one gets :
\begin{equation}
\Delta \phi=\omega\cdot \mathbf B_0
\label{rot}
\end{equation}
where $\omega=\mathbf \nabla \times \mathbf v$ is the local vorticity of the flow \cite{Tsinober}. The measured voltage depends on the vorticity component parallel to the applied magnetic field, i.e. to gradients of the two components of  velocity perpendicular to $\mathbf B_0$. The relation between the flow and the potential is not straightforward but the potential difference can be seen as a linearly filtered measurement of the velocity fluctuations. For length scales larger than the electrode separation, the potential difference can be approximated by the potential gradient, which has the dimension of $vB_0$. The spectral scaling of $\nabla \phi/B_0$ is expected to be the same as that of the velocity, i.e. the $k^{-5/3}$ Kolmogorov scaling. For smaller scales, some filtering results from the finite size of the probe.  For scales smaller than the separation $l$, the values of the potential on the two electrodes are likely to be uncorrelated: if these scales are also in the inertial range, one expects the spectrum of the potential difference to scale as the potential itself, i.e. $k^{-11/3}$ due to the extra spatial derivative. Assuming sweeping of the turbulent fluctuations by the average flow or the energy containing eddies~\cite{Tennekes} and a $k^{-5/3}$ Kolmogorov scaling for the velocity spectrum, we expect the temporal spectrum of the measured potential difference to decay as $f^{-5/3}$ for intermediate frequencies and as $f^{-11/3}$ for high frequencies in the inertial range.

An example of a measured time series of the potential difference is shown in figure~\ref{sigtemp} together with its power spectrum density. For this dataset, one expects a signal of the order of $2\pi Rf_{rot}B_0l\sim 20$~$\mu$V which is the right order of magnitude. Because of the quite large separation of the electrodes, the cutoff frequency between the $f^{-5/3}$ and the $f^{-11/3}$ behaviors is low so that no real scaling is observed but rather trends. At the highest frequencies, the decay is faster than $f^{-11/3}$.

\begin{figure}[!htb]
(a)\includegraphics[width=8cm]{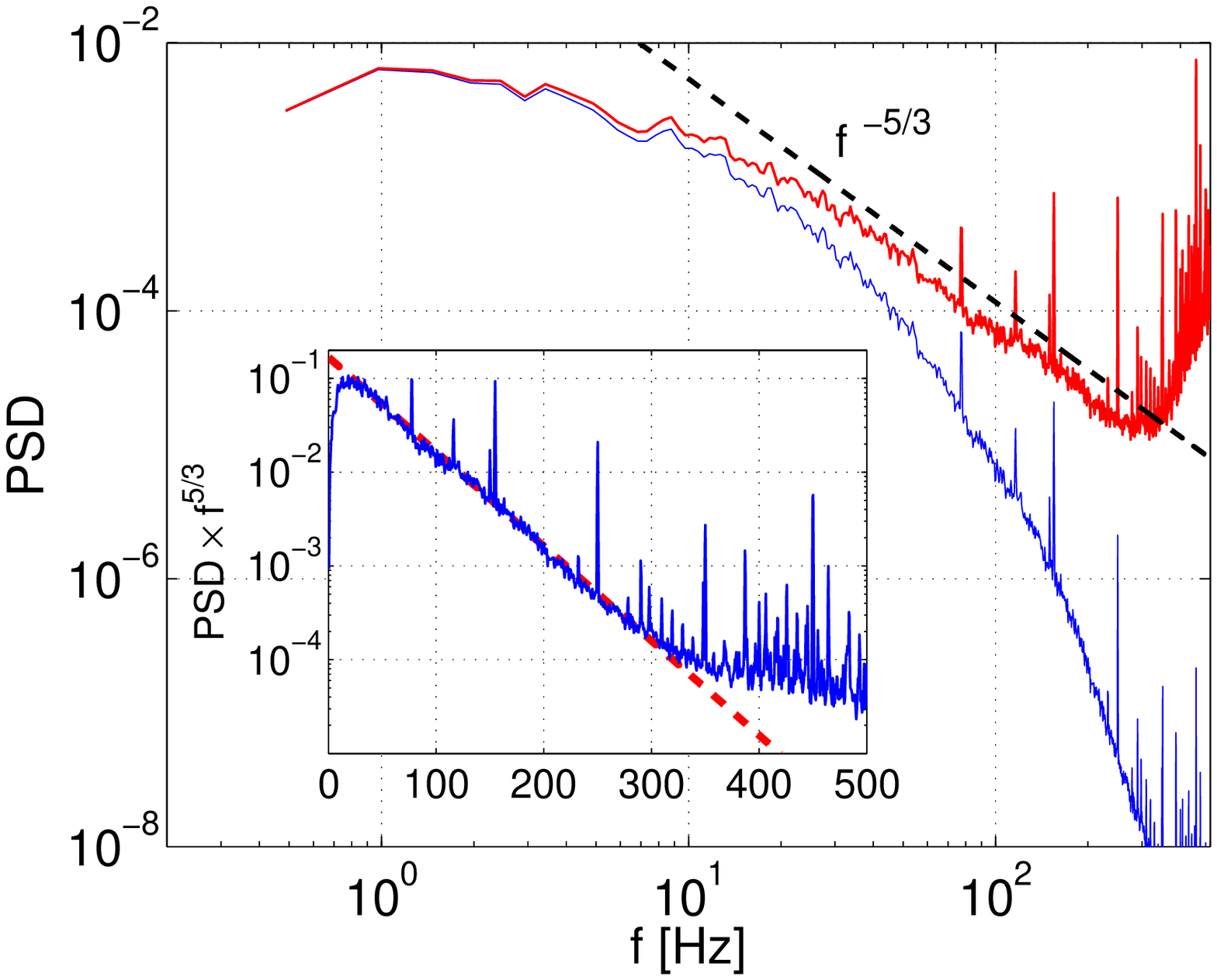}

(b)\includegraphics[width=8cm]{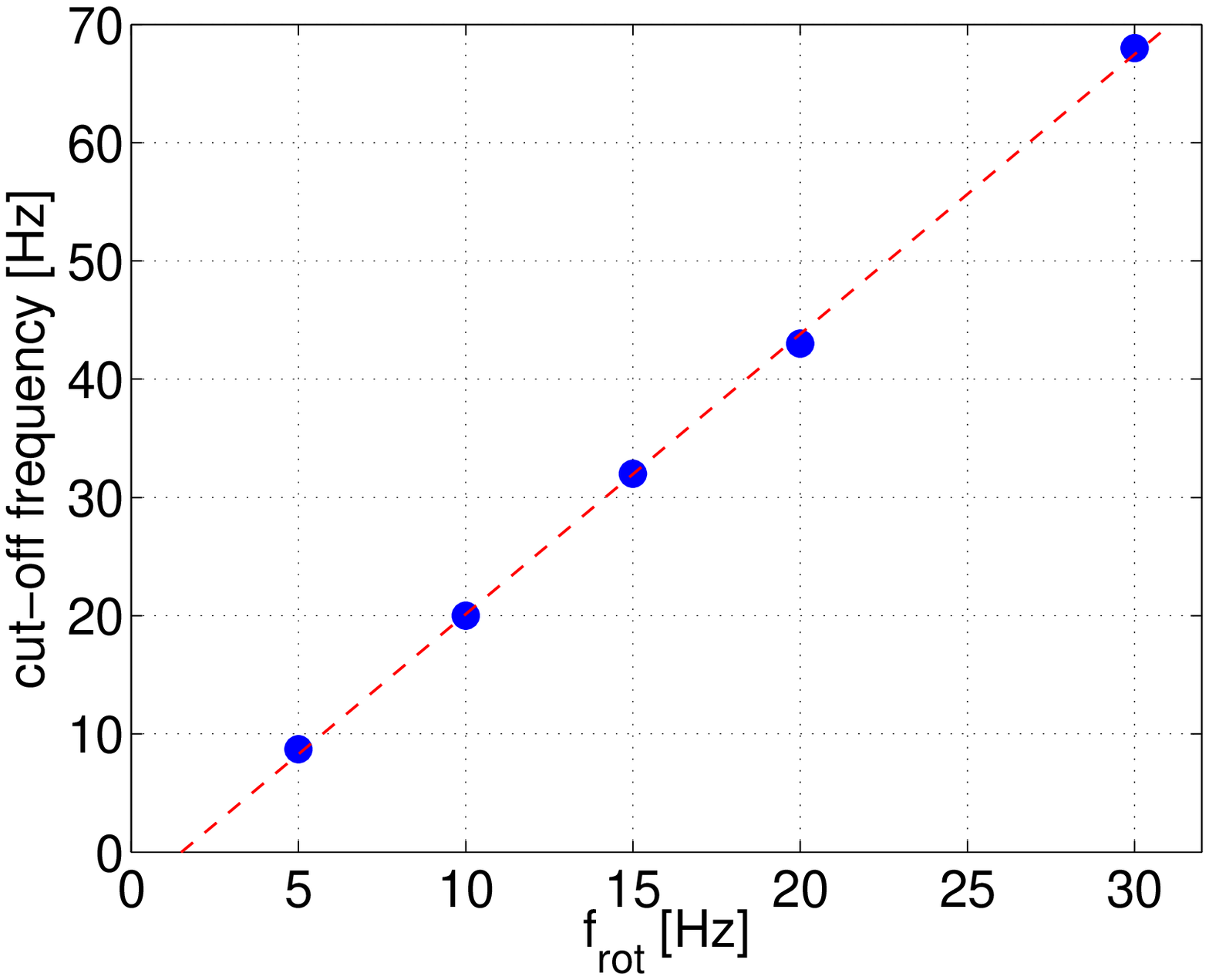}
\caption{\label{compspec}(a) Filtering effect due to the probe geometry. -- Inset: semilog plot of the power spectrum density (PSD) of the potential difference compensated by $f^{5/3}$. The dashed line is an exponential fit. -- Main figure: lower curve, PSD; upper solid line: PSD corrected from the exponential decay fitted in the inset. Dashed line: $f^{-5/3}$ decay. Same data set as that of the previous figure. (b) Cutoff frequency of the probe response for the probe used in (a) and for various rotation frequencies of the propeller. The dashed line is a linear fit. The applied magnetic field is 178~G.}
\end{figure}
Bolonov et al. \cite{Bolonov} had a rather empirical approach to take into account the filtering from the probe. Assuming that the velocity spectrum should decay as $f^{-5/3}$ they observed a spectral response of the potential probe which displays an exponential decay $\exp (-lf/0.6u)$. The factor $0.6$ is most likely dependent on the geometry. We reproduced their analysis in figure~\ref{compspec}. In the inset is displayed the spectrum of figure~\ref{sigtemp} multiplied by the expected $f^{-5/3}$ scaling. The decay is seen to be exponential from about 25~Hz to 300~Hz (at higher frequencies the signal does not overcome the noise). The characteristic frequency of the decay can be extracted and plotted as a function of the rotation frequency of the propeller (the local velocity is expected to scale as $Rf_{rot}$). A clear linear dependence is observed, in agreement with the results of Bolonov {\it et al.}. The cutoff frequency is about twice $f_{rot}$. It is not very high due to the rather large size of the probe. Nevertheless the most energetic length scales are resolved in our measurement. In the following we show only direct spectra, and no correction of the filtering is attempted.

We can conclude that although some filtering is involved, the measurement of potential differences gives an image of the spectral properties of the velocity fluctuations. Any change in the spectral properties of the flow in the vicinity of the probe will thus be visible on the spectrum of the potential difference.

 The potential probes are quite large in order to fit a gaussmeter Hall probe in the vicinity of the electrodes (see Fig. \ref{pot}). These probes are connected to an F.W.Bell Gaussmeter model 7030 that allows to measure the induced magnetic field down to a few tenths of Gauss. The proximity between the velocity and magnetic field measurements allows to study the possible correlations between these two fields.


\section{Effect of the applied magnetic field on the turbulence level}

\subsection{Velocity field}

\begin{figure}[!htb]
\includegraphics[width=8cm]{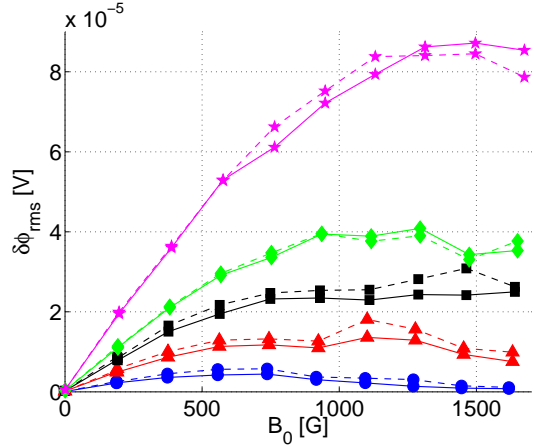}
\caption{\label{phi} Evolution of the rms value of the potential difference as a function of the applied vertical magnetic field, for different values of the rotation frequency. $\bullet$: $f_{rot}=5$~Hz, $\triangle$: 10~Hz, $\square$: 15~Hz, $\diamond$: 20~Hz and $\star$: 30~Hz. The solid lines correspond to the azimuthal potential difference and the dashed line to the radial potential difference for the same four electrode potential probe.}
\end{figure}
We first focus on the fluctuation level of the velocity field accessed through measurements of potential differences. The evolution of the {\it rms} value of the potential difference is displayed in figure~\ref{phi} as a function of the applied magnetic field and of the rotation rate. From equation (\ref{pot}), the potential difference should behave as $\delta\phi_{rms}\propto B_0v/l$. For low values of the applied magnetic field, the interaction parameter $N$ is low: the magnetic field has almost no effect on the flow, and the velocity scales as $Rf_{rot}$. Thus, for high $f_{rot}$ and low $B_0$ (i.e. low $N$), $\delta \phi_{rms}$ sould be linear in both $f_{rot}$ and $B_0$. The upper curve corresponds to the highest velocity $f_{rot}=30$~Hz. For low $B_0$ there is a linear increase of $\delta\phi_{rms}$. Then it seems to saturate. For smaller rotation rates, the linear part gets smaller and the saturation region gets wider. Eventually, for $f_{rot}=5$~Hz, the potential decays for the highest values of the applied magnetic field. This demonstrates that there is a strong interaction between the magnetic field and the flow.

\begin{figure}[!htb]
\includegraphics[width=8cm]{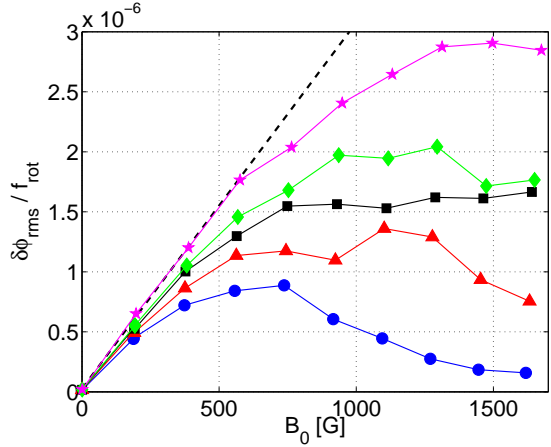}
\caption{\label{phiofrot} Evolution of the rms value of the azimuthal potential difference normalized by the rotation frequency, as a function of the applied vertical magnetic field. $\bullet$: $f_{rot}=5$~Hz, $\triangle$: 10~Hz, $\square$: 15~Hz, $\diamond$: 20~Hz and $\star$: 30~Hz. The dashed line is a linear trend fitted on the first four points of the 30 Hz data.}
\end{figure}
To investigate in more details the scaling properties of the potential difference, $\delta\phi_{rms}/f_{rot}$ is displayed versus $B_0$ in figure~\ref{phiofrot}. One can clearly see that as $f_{rot}$ decreases, the potential deviates from the linear trend for lower and lower values of the magnetic field.

\begin{figure}[!htb]
\includegraphics[width=8cm]{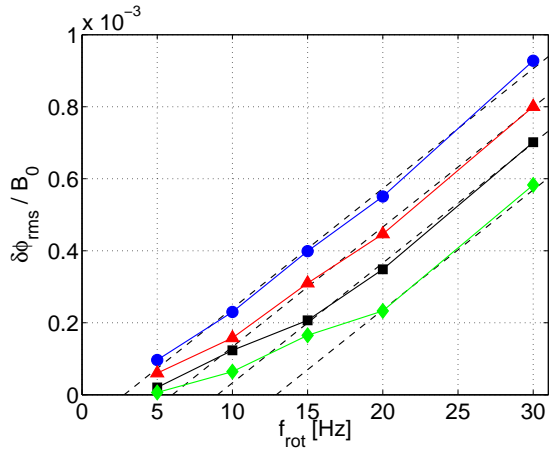}
\caption{\label{phioB0} Evolution of the rms value of the azimuthal potential difference normalized by the imposed vertical magnetic field, as a function of the rotation frequency of the propeller. $\bullet$: $B_0=356$~G, $\triangle$: 712~G, $\square$: 1070~G, $\diamond$: 1420~G. The parallel dashed lines are used as eye guides. The upper one corresponds to a linear fit of the data at $B_0=356$~G.}
\end{figure}
Figure~\ref{phioB0} shows $\delta\phi_{rms}/B_0$ as a function of $f_{rot}$. Here a linear trend is observed for large $f_{rot}$. As the external magnetic field is increased, the fluctuations of the potential are damped and the linear trend is recovered for increasingly high values of the rotation rate. 

\begin{figure}[!htb]
(a)\includegraphics[width=8cm]{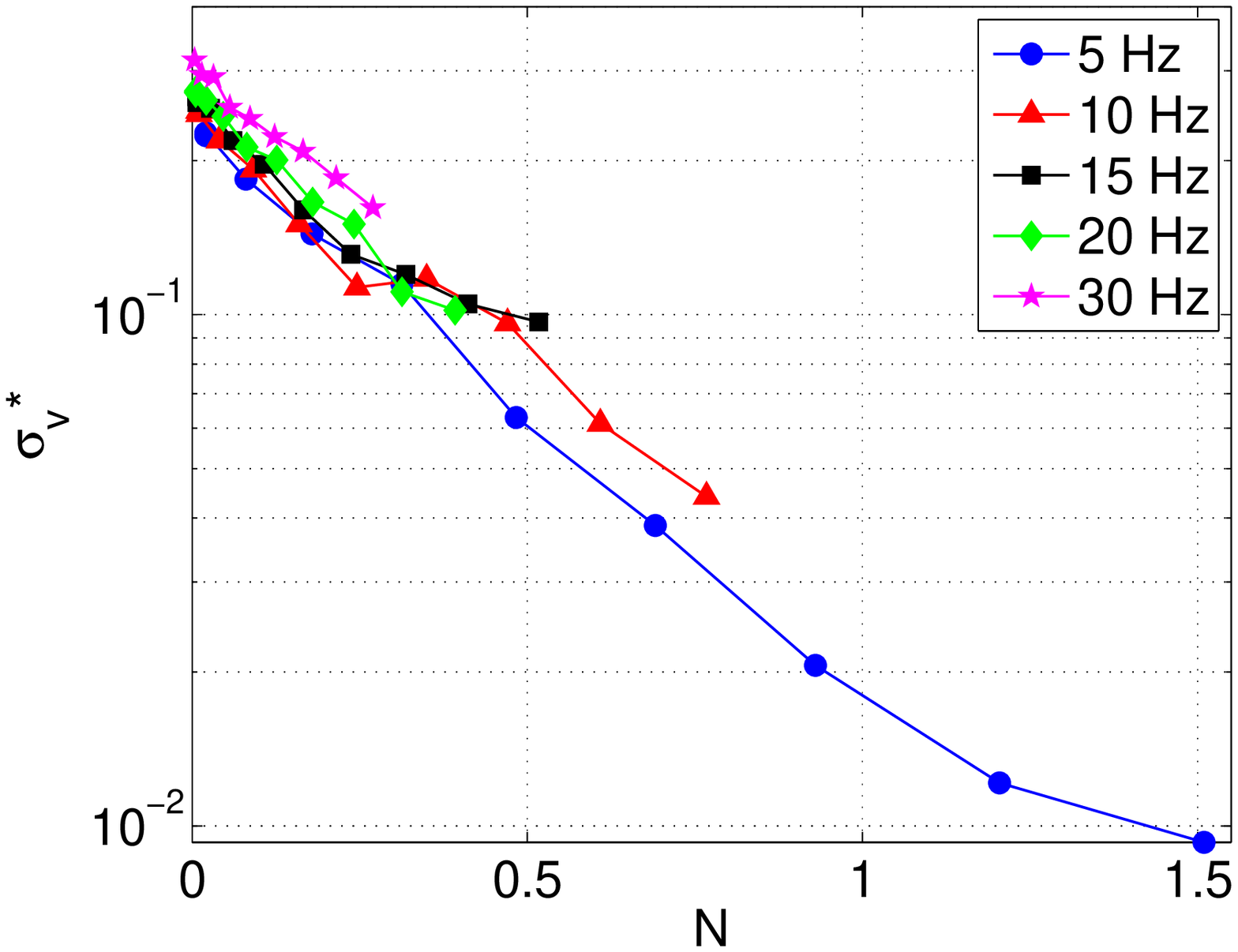}
(b)\includegraphics[width=8cm]{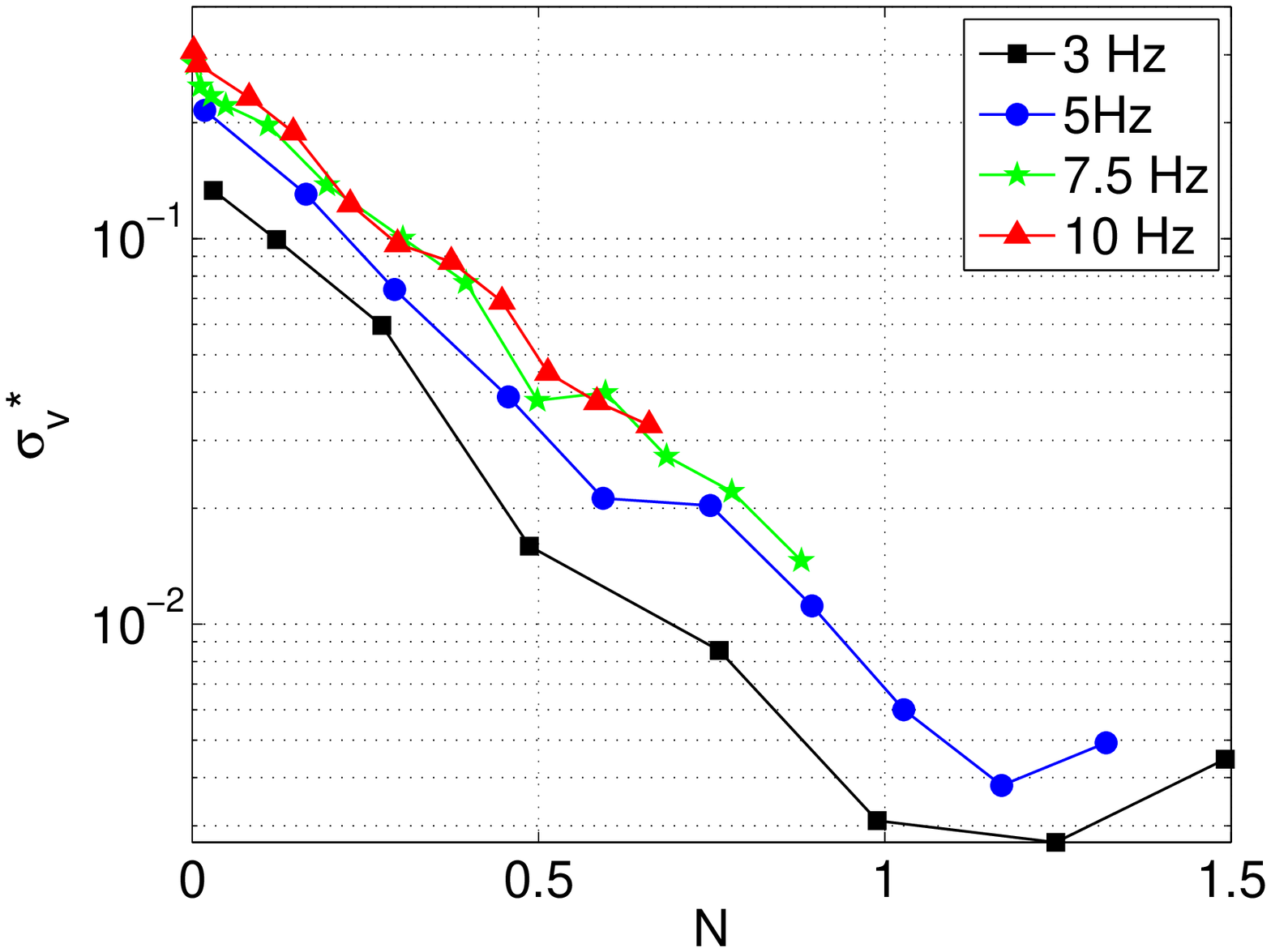}
\caption{\label{sigvstar} Evolution of the dimensionless potential $\sigma_v^\star$ (see text) as a function of the interaction parameter $N$. (a) and (b) correspond to the azimuthal potential difference taken from two different datasets. The potential probes are similar but not perfectly identical. The rotation frequencies reach the highest values in (a) and the lowest in (b). The representation is semilogarithmic.}
\end{figure}
All that information can be synthesized by plotting the dimensionless potential
\begin{equation}
\sigma_v^\star=\frac{\delta\phi_{rms}}{B_0lRf_{rot}}\, .
\end{equation}
This quantity can be understood as the velocity fluctuations of the flow normalized by the forcing velocity of the propeller. This quantity is displayed as a function of the interaction parameter $N=\frac{\sigma L B_0^2}{2\pi \rho R f_{rot}}$ in figure~\ref{sigvstar}. In this representation, the data collapses fairly well on a single master curve for high rotation rates. The damping of the turbulent fluctuations can reach one order of magnitude for $N$ close to 1. In fig.~\ref{sigvstar}(b) - that corresponds to lower rotation rates - there is a slight and systematic drift of the curves  with the rotation rate. This indicates a slight dependence in Reynolds number that comes likely from the fact that the flow is not fully similar with $f_{rot}$ for low values of this rotation rate. The main dependence is clearly in the interaction parameter. The velocity fluctuations are seen to decay exponentially with $N$ for values of $N$ up to order 1. For higher $N$ the data is affected by the noise. The velocity fluctuations are so strongly damped that the signal to noise ratio decreases significantly, as can be seen on the spectra in the following sections. The decay rate is about $2.5$ in fig.~\ref{sigvstar}(a) and $3.5$ in (b). The difference may come from geometrical factors of the probes which are not exactly similar in both datasets, or from a slight mismatch between the positions of the probes in the two datasets. We observed a similar collapse of potential data with $N$ in a different experiment~\cite{Berhanu}. This older experiment was smaller in size, with only one propeller and smaller magnetic field. Only the beginning of the exponential decay could be observed in that case.

\subsection{Induced magnetic field}
\label{indmag}

\begin{figure}[!htb]
(a)\includegraphics[width=8cm]{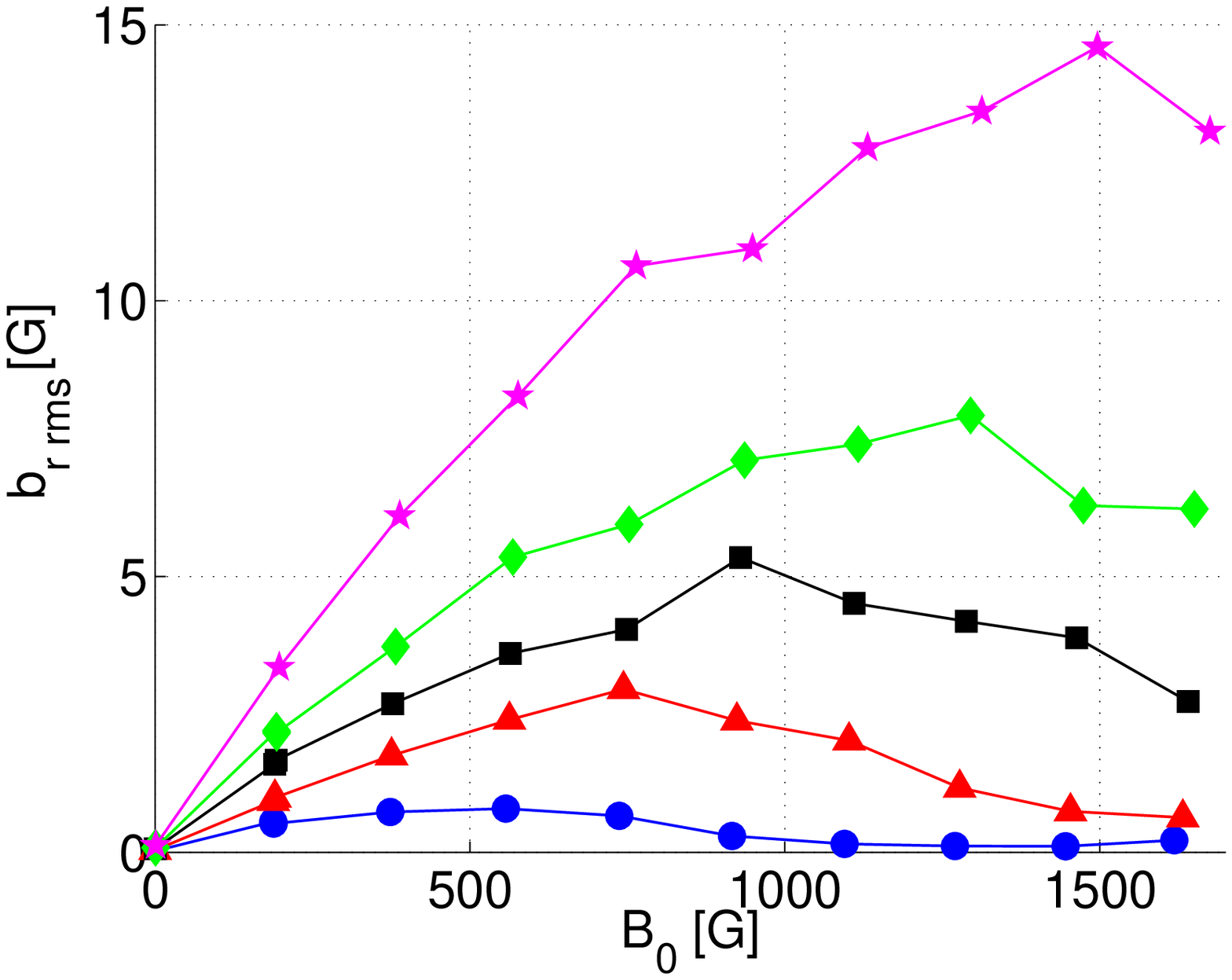}

(b)\includegraphics[width=8cm]{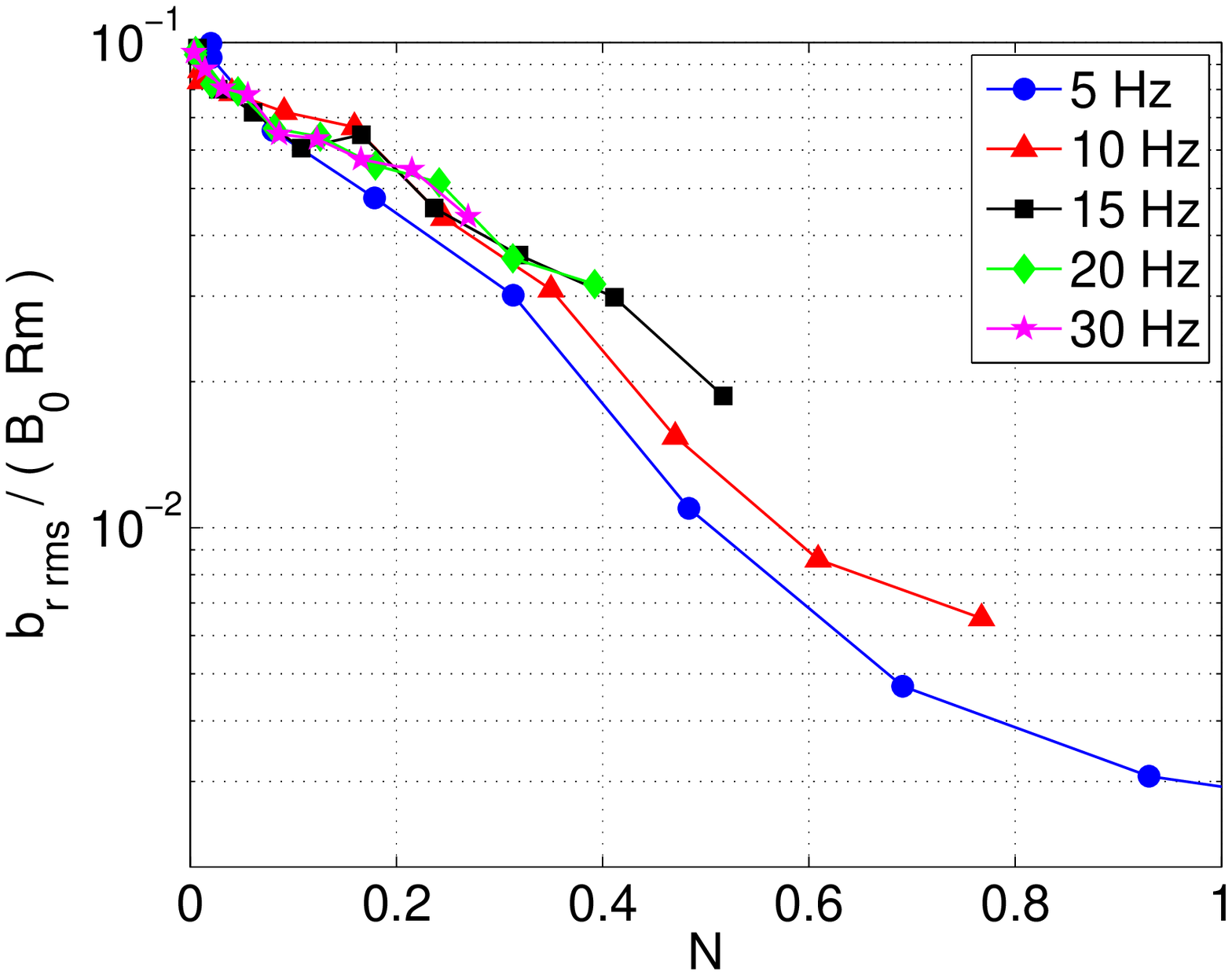}

\caption{\label{b} (a) Evolution of the rms value of the induced radial magnetic field as a function of the applied vertical magnetic field, for different values of the rotation frequency. $\bullet$: $f_{rot}=5$~Hz, $\triangle$: 10~Hz, $\square$: 15~Hz, $\diamond$: 20~Hz and $\star$: 30~Hz. (b)~Evolution of ${b_{r rms}}/{B_0Rm}$ as a function of the interaction parameter $N$. The representation is semilogarithmic.}
\end{figure}
The induced magnetic field is of the order of one percent of the applied field. This low value is due to the low magnetic Reynolds number which is at best of order 1. For low $Rm$ the induction equation reduces to 
\begin{equation}
\mathbf B_0\cdot\nabla \mathbf v+\eta\Delta\mathbf b=0
\label{lowRm}
\end{equation}
(assuming that $\mathbf B_0$ is uniform). The induced magnetic field thus reflects the velocity gradients in the direction of the applied magnetic field. We use it as a second tool to investigate the statistical properties of the flow.

The fluctuation level of the radial magnetic field $b_{r rms}$ is displayed in figure~\ref{b}(a). Its evolution with $B_0$ and $f_{rot}$ is strongly similar to that of the potential differences, as expected from the previous arguments. The azimuthal component displays the same behavior (not shown).

For low values of the applied magnetic field, the flow is not affected very much by the Lorentz force and from equation (\ref{lowRm}) one expects the induced magnetic field amplitude to scale as: $b\propto B_0Rm$ (see \cite{Odier} for example). From the previous section, the amplitude of the velocity field decays exponentially with $N$. We thus expect $b_{rms}/B_0Rm$ to have the same qualitative behavior. This is what is observed in figure~\ref{b}(b). The various datasets are seen to collapse on a single master curve for $N$ up to 0.5. For higher $N$, the sensitivity of our gaussmeter is not high enough for the signal to overcome the electronic noise. As for the velocity, the curve for the lowest $f_{rot}$ is below all the others, which confirms the slight dependence with $Rm$ observed on the potential measurements. At a given value of $N$, the collapse of the measurements shows that the fluctuations of the induced magnetic field are indeed linear both in applied magnetic field and in magnetic Reynolds number: $b_{rms}\simeq 0.1 B_0Rm$. 

\begin{figure}[!htb]
(a)\includegraphics[width=8cm]{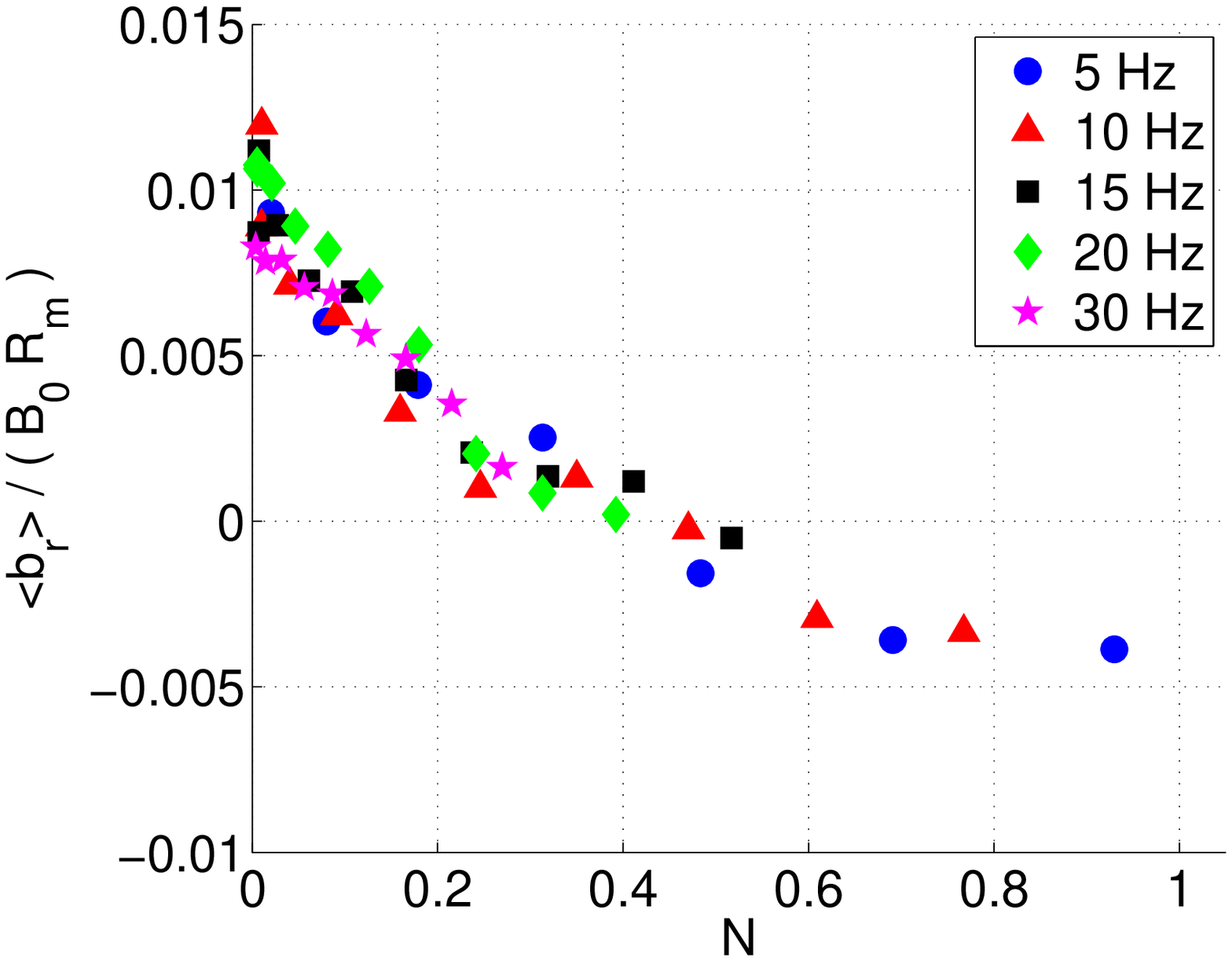}
(b)\includegraphics[width=8cm]{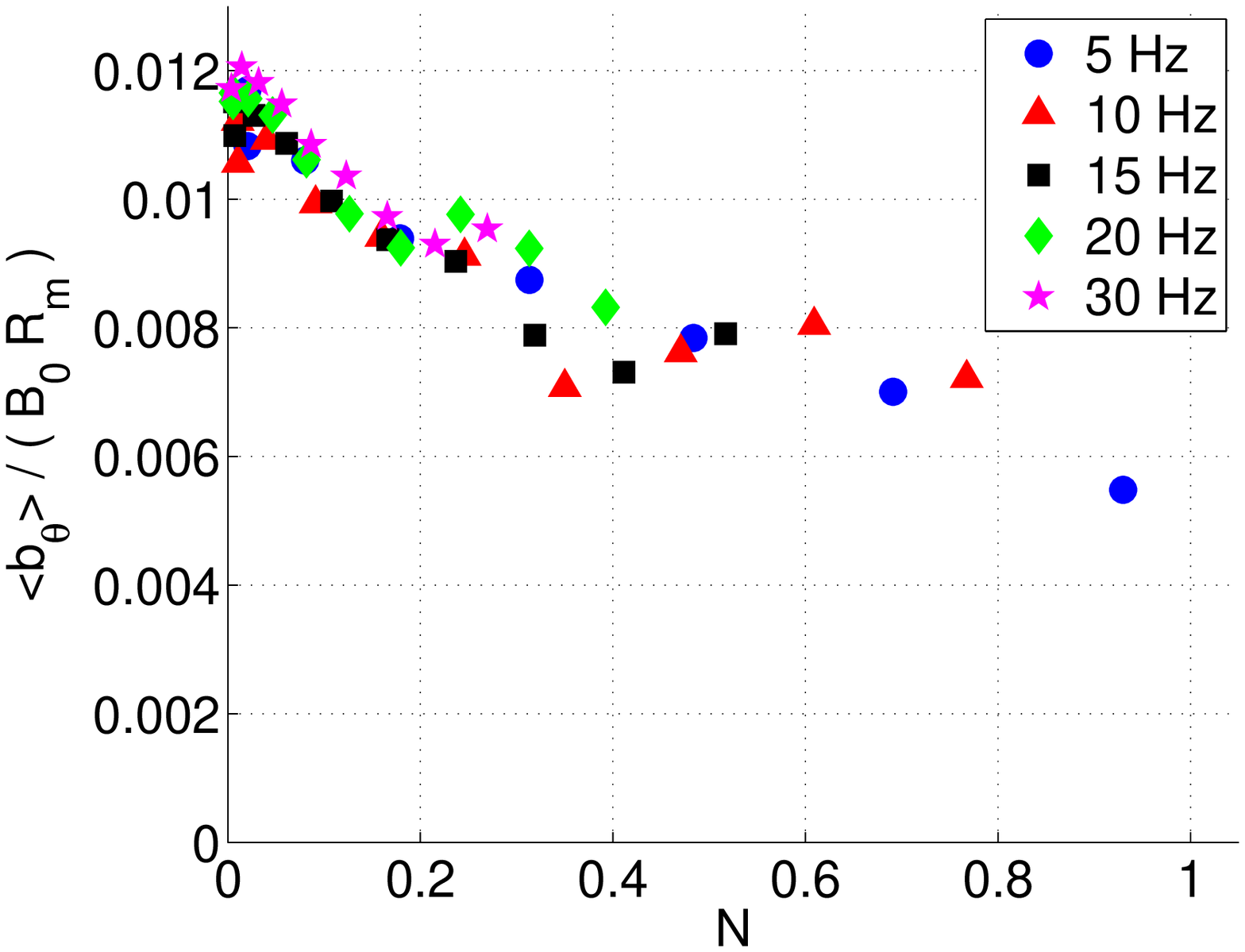}
\caption{\label{moyb} (a) Evolution of $\dfrac{\langle b_r\rangle}{B_0Rm}$ as a function of the interaction parameter $N$. (b) same for $\dfrac{\langle b_\theta\rangle}{B_0Rm}$.}
\end{figure}
We have also measured the average value of the induced magnetic field. From equation (\ref{lowRm}) it is linked to the vertical gradients of the time-averaged velocity field. $\langle b_r\rangle/B_0Rm$ and $\langle b_\theta\rangle/B_0Rm$ are shown in figure~\ref{moyb} as a function of $N$. For both components, this representation leads to a good collapse of the datasets. At a given value of $N$, the collapse shows again that the average induced magnetic field is linear in $Rm$ and $B_0$ with $\langle b_i \rangle \simeq 0.01 B_0 Rm$ for the data shown here. In an ideal Von K\'arm\'an experiment, when the two propellers counter-rotate at the same speed, the time averaged flow is invariant to a rotation of angle $\pi$ around a radial unit vector taken in the equatorial plane (denoted as $\vec{e}_r$ on figure \ref{setup}). If the applied field were perfectly symmetric and the probe positioned exactly in the equatorial plane, this symmetry should lead to $\langle b_r\rangle=0$. However, the introduction of the probe breaks the symmetry and neither the mecanical device nor the applied magnetic field are perfectly symmetric. It has been observed in our setup and in the VKS experiment that the measurement of $\langle b_r\rangle$ is extremely sensitive to the position of the probe (private communication from the VKS collaboration). For example, a slight mismatch of the propeller frequencies or of the relative positions of the probe and mid-plane shear layer can lead to strong changes in the mean value of the magnetic field. This may be the reason why $\langle b_r\rangle$ is not zero here. Nevertheless there is a systematic change of the average with $N$ which is consistent across the different values of the velocity and of the applied magnetic field. It indicates a change of the time-averaged flow in the vicinity of the probe. To the shear layer lying in the equatorial plane corresponds strong radial vorticity, orthogonal to the applied magnetic field. The applied strong magnetic field will impact the shear layer, as it tends to elongate the flow structures along its axis. Even a small change in the shear layer geometry affects strongly the measured $\langle b_r\rangle$. Here we see that its sign is changed at high $N$. 

The time-averaged induced azimuthal magnetic field $\langle b_\theta \rangle$ is due to $\omega$-effect from the differential rotation of the propellers \cite{Moffatt, Odier}. It is seen to decay slightly (about 35$\%$) with $N$. This effect may be due to some magnetic braking that leads to an elongation of the shear layer and thus to weaker differential rotation in the vicinity of the mid-plane. Nevertheless this effect is limited and we expect the average large scale structure of the flow to remain almost unchanged. The small change of the large scale flow is not the reason for the strong damping of the turbulent fluctuations by an order of magnitude.

\subsection{Injected mechanical power}

\begin{figure}[!htb]
\includegraphics[width=8cm]{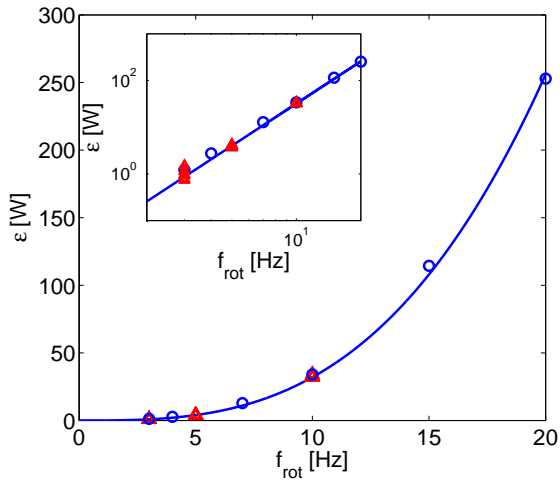}
\caption{\label{power} Injected mechanical power as a function of the rotation frequency of the propellers. The circles correspond to measurements without magnetic field, and the triangles correspond to measurements at constant rotation frequency and $B_0=90, 180, 530, 705, 880, 1000, 1230, 1240, 1320, 1410$, and $1500$ Gauss. The solid line has equation $y=0.032 f_{rot}^3$  which comes from the turbulent scaling law. The inset is a log-log representation.}
\end{figure}

An interesting issue in MHD turbulence is to understand how the injected mechanical power is shared between ohmic and viscous dissipations. The torques $T_1$ and $T_2$ provided by the motors can be accessed through measurements of the current delivered to them. The injected mechanical power is then $\epsilon=(T_1+T_2) 2 \pi f_{rot}$. This quantity has been measured as a function of the rotation frequency with and without applied magnetic field. The results are drawn on figure \ref{power}: without magnetic field, the injected power follows the turbulent scaling law $\epsilon \sim f_{rot}^3$. More surprisingly, we notice that at a given rotation frequency the experimental points corresponding to different amplitudes of the magnetic field are indistinguishable. This observation confirms that no change of the global structure of the flow is induced by the magnetic field. The fact that the injected power is independent of the applied magnetic field seems to be in contradiction with results from numerical simulations where an ohmic dissipation of the same order of magnitude as the viscous one is observed when a strong magnetic field is applied (the ohmic dissipation is around three times the viscous dissipation for $N=1$ in Burattini et al.\cite{Burattini}). One could argue that there may be a large ohmic dissipation compensated by a drop in the viscous dissipation: the energy flux in the turbulent cascade would then be dissipated mainly through ohmic effect without changing the overall injected power. However, a rough estimate of the ohmic dissipation $D_j$ gives values which are rather low: with the maximum value of the induced magnetic field $b \simeq 0.1 B_0 Rm$, and assuming that the magnetic field is dissipated mostly at large scale, one gets:
\begin{equation}
D_j \sim \frac{j^2}{\sigma} L^3 \sim \frac{1}{\sigma} \frac{(0.1 Rm B_0)^2}{L^2 \mu_0^2} L^3 \simeq 4 W
\end{equation}
where we used the values $Rm=1$, and $B_0=1500 G$. This estimate is much lower than the injected power and we may expect ohmic dissipation to remain negligible compared to viscous dissipation, even at order one interaction parameter. However, one also needs to know the current that leaks through the boundaries to evaluate the additional ohmic dissipation that takes place inside the walls. As these effects are difficult to quantify, we are not able to determine the ratio of ohmic to viscous dissipation. Nevertheless it is interesting to notice that the injected mechanical power remains the same when a strong magnetic field is applied, although velocity fluctuations are decreased by a factor 10 in the mid-plane of the tank.

\subsection{Development of anisotropy}

\begin{figure}[!htb]
\includegraphics[width=8cm]{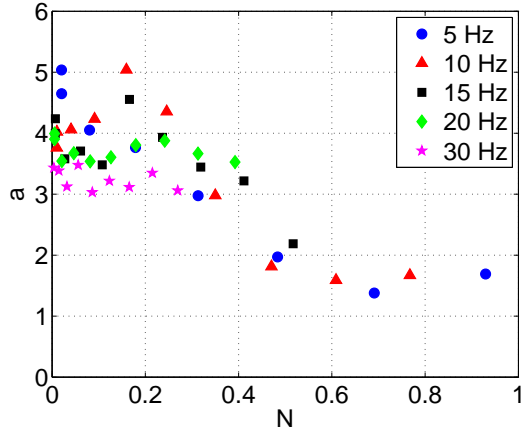}
\caption{\label{anis} Evolution of the anisotropy parameter $a=\eta \frac{b_{r rms}}{\delta\phi_{rms}}$ as a function of the interaction parameter.}
\end{figure}

The application of a uniform magnetic field on a turbulent flow is known to elongate the flow structures in the direction of the applied field \cite{Knaepen}. In decaying turbulent flows, this effect eventually leads to the bidimensionalization of the flow \cite{Sommeria}. As far as forced turbulence is concerned, only numerical simulations have demonstrated this effect \cite{Vorobev}. The measurement of both the induced magnetic field and the electric potential allows to quantify the elongation of the turbulent structures in the $z$ direction: on the one hand, equation~(\ref{rot}) links the electric potential to the vertical component of vorticity, i.e. to horizontal gradients of velocity. On the other hand the induced magnetic field is related to vertical gradients of velocity through equation (\ref{lowRm}). We expect then that
\begin{equation}
\frac{\Delta b}{\Delta \phi} \sim \frac{1}{\eta} \frac{\partial_{||} v}{\partial_\perp v}
\end{equation}
where $\partial_{||}$ and $\partial_\perp$ denote derivatives in directions parallel and perpendicular to the applied magnetic field.
As we cannot access experimentally the order of magnitude of the Laplacians, we define the quantity $a=\eta \frac{b_{r rms}}{\delta\phi_{rms}}$ which we expect to give a crude estimate of the ratio of the vertical to the horizontal gradients of velocity. It is somewhat related to the parameter $G_1$ defined in Vorobev et al.~\cite{Vorobev}.
This anisotropy parameter $a$ is represented as a function of the interaction parameter $N$ in fig.~\ref{anis} for different values of the rotation frequency: it decreases from about 4 until it saturates around 1.5. This decrease of the parameter $a$ by a factor about 3 when a strong magnetic field is applied provides evidence for the elongation of the flow structures in the $z$ direction: the derivatives of the velocity field in the direction of $\mathbf B_0$ are much smaller than its derivatives in directions perpendicular to $\mathbf B_0$. Although the turbulence becomes more anisotropic, it remains three dimensional even for the highest value of the interaction parameter reached in this experiment. This is due to the fact that the forcing imposed by the propellers is three dimensional and prevents the flow from becoming purely two dimensional.

\section{Temporal dynamics}

We observe that the turbulent fluctuations are being damped by magnetic braking when a strong magnetic field is applied to homogeneous and nearly isotropic turbulence. An interesting question is to know how this damping is shared among scales. In this section we study the evolution of the power spectrum densities of the potential difference and induced magnetic field.

\subsection{Potential}

\begin{figure}[!htb]
(a)\includegraphics[width=7cm]{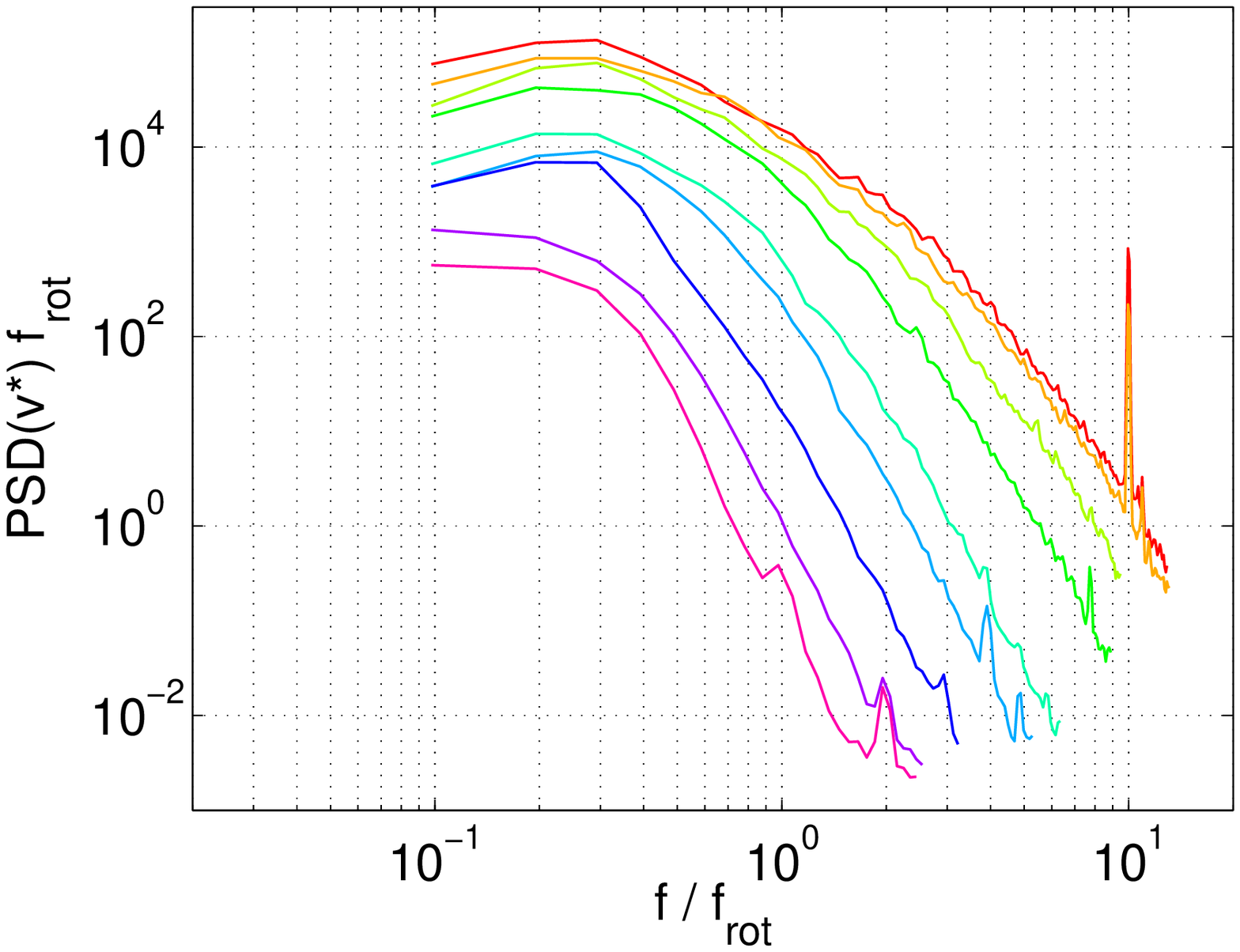}
(b)\includegraphics[width=7cm]{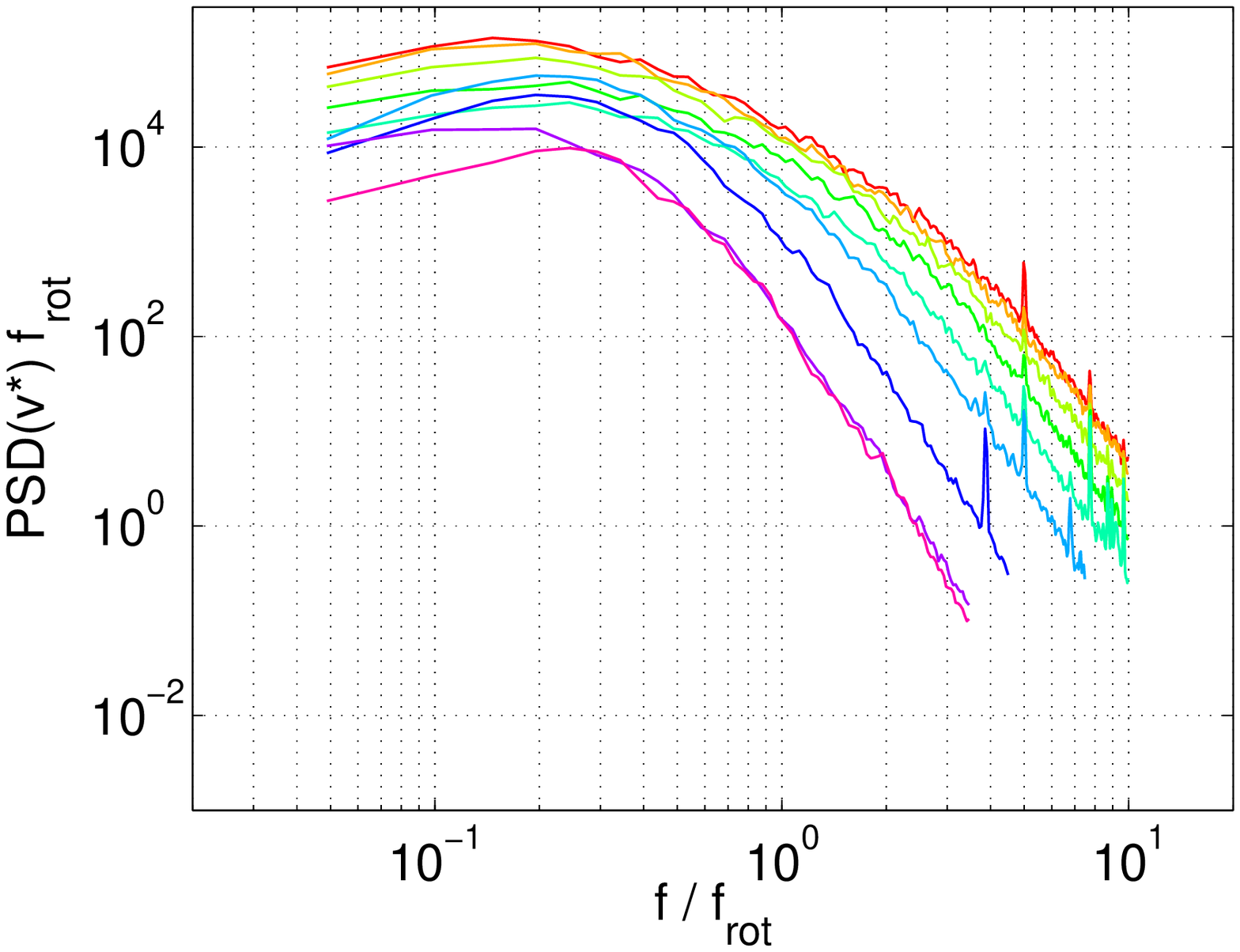}
(c)\includegraphics[width=7cm]{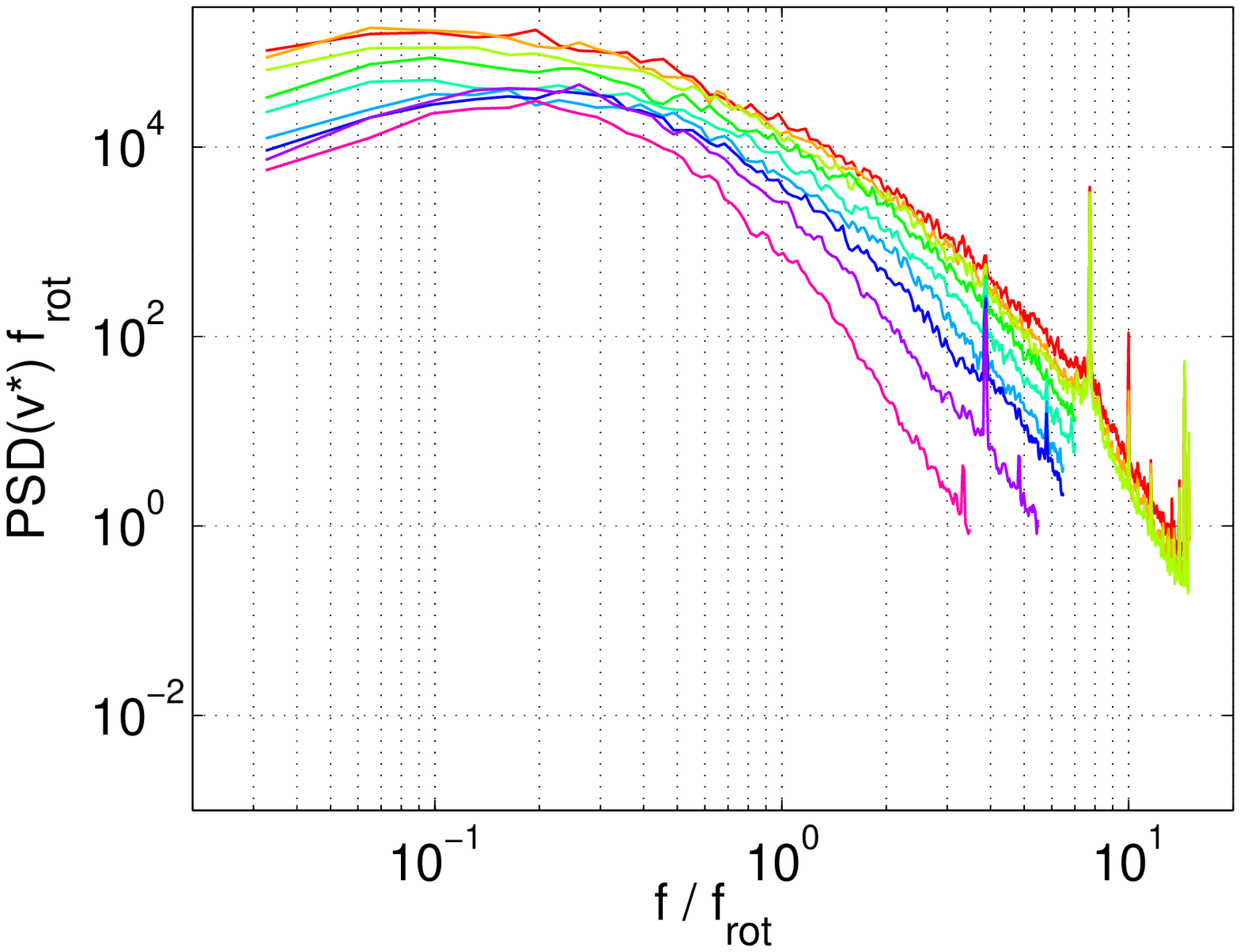}
\caption{\label{spv} Evolution of the power spectrum density of the dimensionless potential $v^\star=\frac{\delta \phi}{B_0 l R f_{rot}}$ for the radial potential difference. The subfigures correspond to different rotation rates of the propellers: (a) $f_{rot}=5$~Hz, (b) 10~Hz, (c) 15~Hz.
In each subfigure, the different curves correspond to the different values of the applied magnetic field $B_0=178$, 356, 534, 712, 890, 1070, 1250, 1420 and 1600~G. They are naturally ordered from top to bottom as the magnetic field is increased. The noise part of the spectra has been removed to improve the clarity of the figures.}
\end{figure}
We show in figure~\ref{spv} the power spectrum densities (PSD) of the dimensionless potential $v^\star={\phi}/{B_0 l R f_{rot}}$. For a rotation frequency of 20~Hz, the spectra are seen to decay as the magnetic field is increased, the shape of the different spectra remaining the same (fig.~\ref{spv}(c)). The decay of the PSD is at most of a factor 10. This dataset reaches a maximum value of the interaction parameter about 0.4. For the lowest displayed value of the rotation frequency $f_{rot}=5$~Hz (fig.~\ref{spv}(a)), $N$ reaches 1.5. As the interaction parameter increases, two distinct regimes are observed: first an overall decay of the PSD and second a change in the shape of the PSD. The highest frequencies are overdamped compared to the lowest ones: for the highest value of $N$, the PSD decays by about 6 orders of magnitude at twice the rotation frequency, whereas it decays by only two orders of magnitude at low frequency. We already observed the first regime in a previous experiment performed on a different flow~\cite{Berhanu}. The interaction parameter was not high enough in this experiment to observe the second regime.

\begin{figure}[!htb]
(a)\includegraphics[width=7cm]{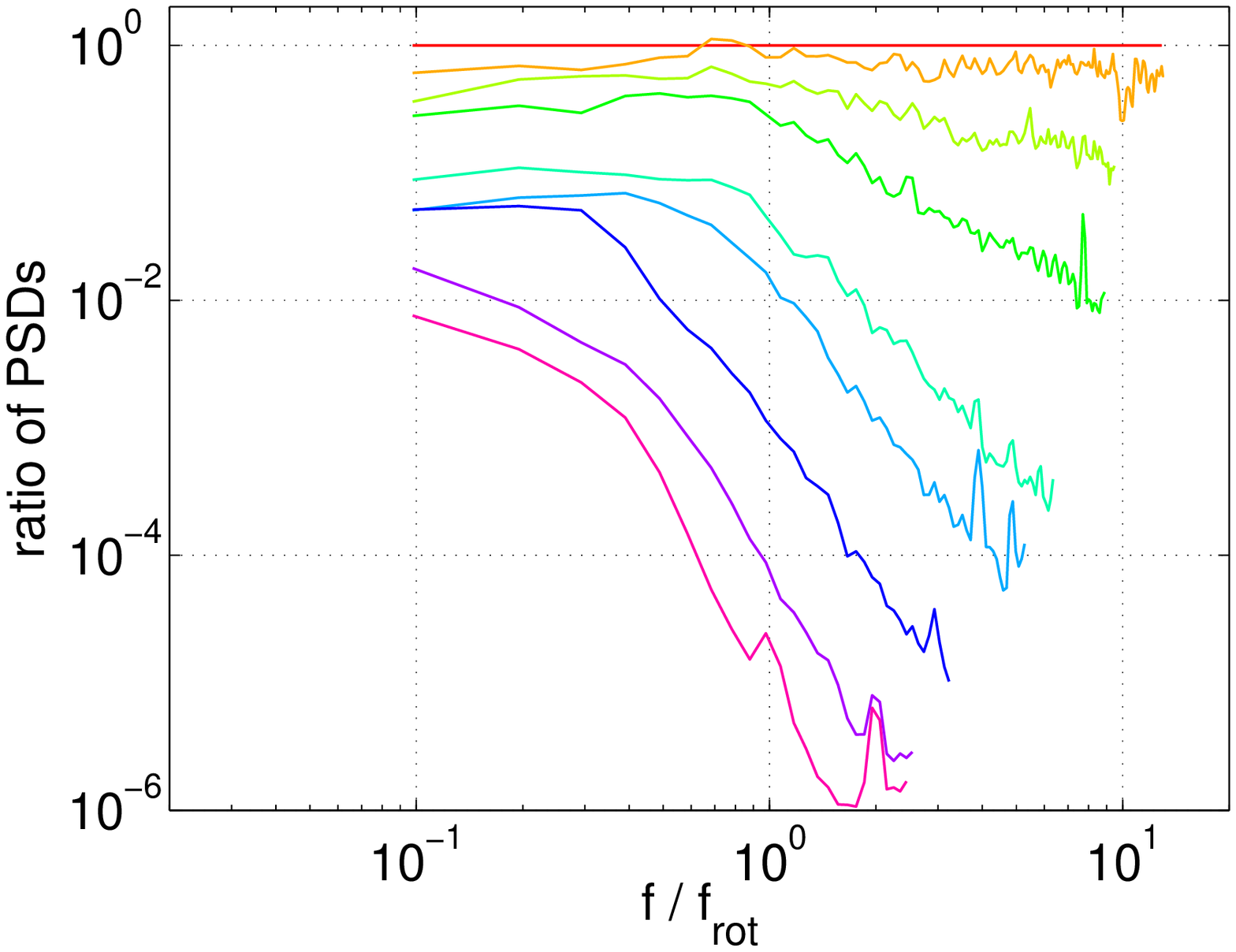}
(b)\includegraphics[width=7cm]{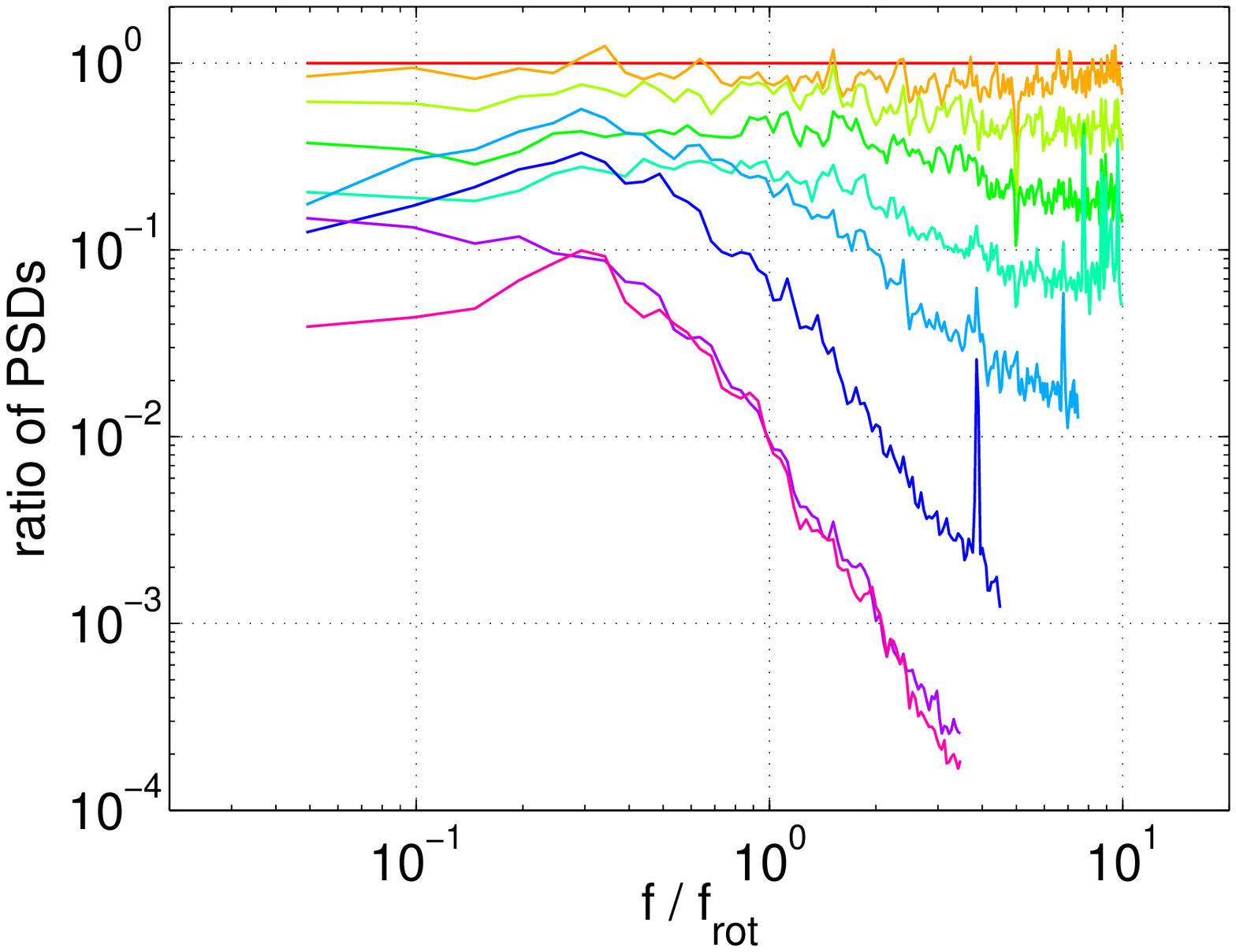}
(c)\includegraphics[width=7cm]{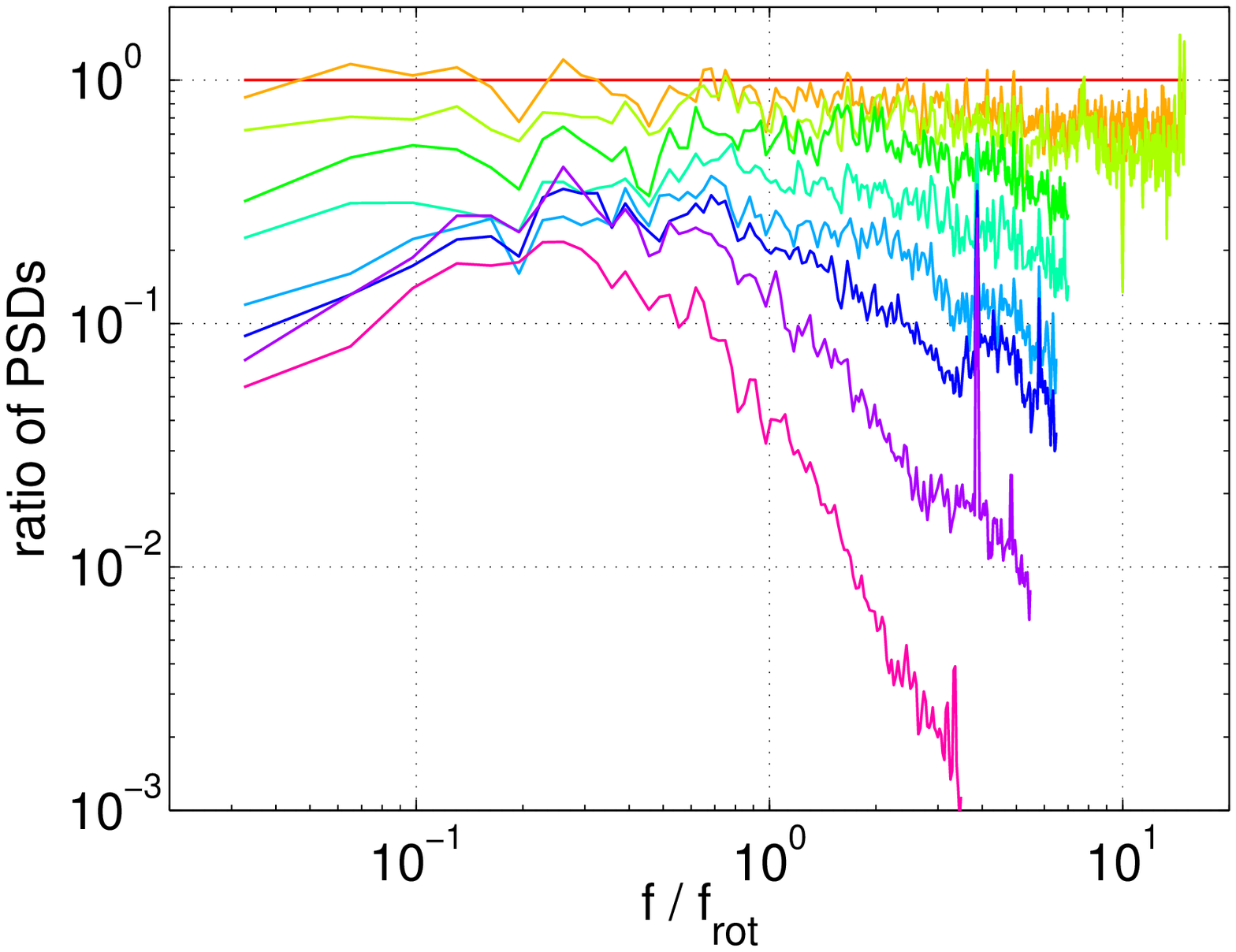}
\caption{\label{spvn} Evolution of the shape of the dimensionless potential's power spectrum. The spectra of figure~\ref{spv} are divided by the spectrum obtained at the smallest value of the applied magnetic field $B_0=178$~G and at the same $f_{rot}$. The subfigures correspond to different rotation rates of the propellers: (a) $f_{rot}=5$~Hz, (b) 10~Hz, (c) 15~Hz.}
\end{figure}
To quantify more precisely the relative decay, we plot in figure~\ref{spvn} the ratio of the PSD of $v^\star$ over the PSD at $B_0=178$~G (and at the same rotation rate of the propeller). At the smallest values of $N$, the ratio weakly changes across the frequencies but it decays with $N$. When $N$ is increased over approximately $0.1$ an overdamping is observed at the highest frequencies. This extra damping can be qualitatively characterized by a cutoff frequency, which decreases  extremely rapidly and seems to reach the rotation frequency for $N \simeq 0.3$. Above the cutoff frequency, the decay rate seems to behave as a power law of the frequency. The exponent of this power law gets more and more negative as $N$ increases and seems to reach $-5$ at the highest value of $N$ displayed here ($f_{rot}=5$~Hz, $B_0=1600$~G, $N=1.5$). 

\begin{figure}[!htb]
\includegraphics[width=9.5cm]{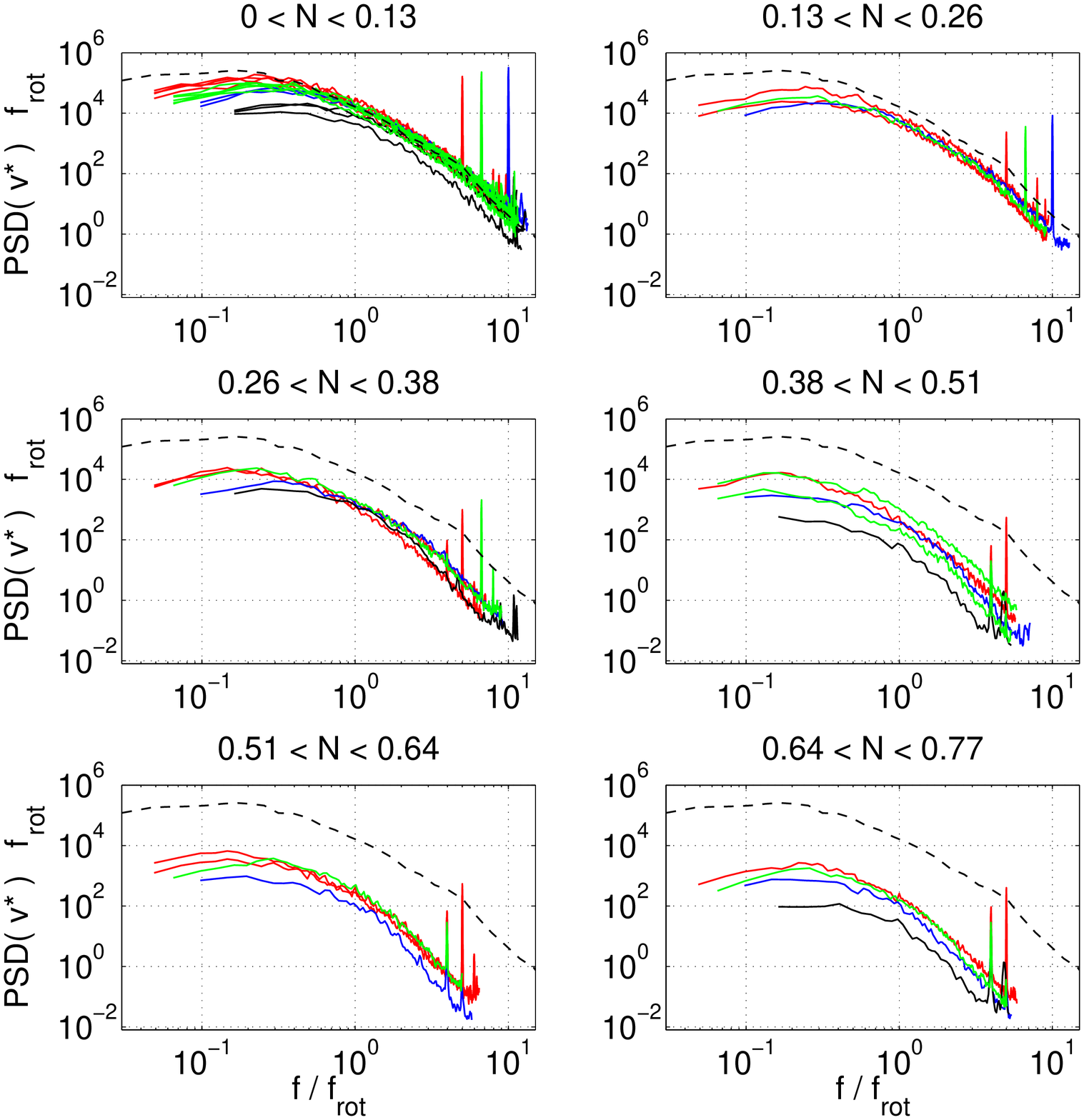}
\caption{\label{spN} Evolution of the shape of the normalized potential's power spectrum as a function of the interaction parameter. Colors corresponds to the different rotation speeds: black 3~Hz, blue 5~Hz, green 7.5~Hz and red 10~Hz. Each subfigure corresponds to data restricted to the specified interval of $N$. The dashed line corresponds to the spectrum at the smallest non zero value of $N$.}
\end{figure}
In figure~\ref{spN}, we gathered the dimensionless spectra at various rotation rates that correspond to the same interval of $N$. The spectra are collapsing fairly well onto each other. A little bit of scatter is observed, most likely due to the slight dependence on $Rm$ described previously. The shape of the spectra is essentially a function of the interaction parameter.

In experiments on the influence of a magnetic field on decaying turbulence, the velocity spectrum goes from an $f^{-5/3}$ to an $f^{-3}$ behavior as the interaction parameter is increased. This $-3$ exponent is attributed either to two-dimensional turbulence or to a quasi-steady equilibrium between velocity transfer and ohmic dissipation. As far as forced turbulence is concerned, we observe a strong steepening of the velocity spectrum, with slopes already much steeper than $f^{-3}$ for $N=1$. We do not observe any signature of this quasi-steady equilibrium or of 2D turbulence. Once again, this comes from the three-dimensional forcing of the propellers which prevents the flow from becoming purely 2D.

\begin{figure}[!htb]
\includegraphics[width=8cm]{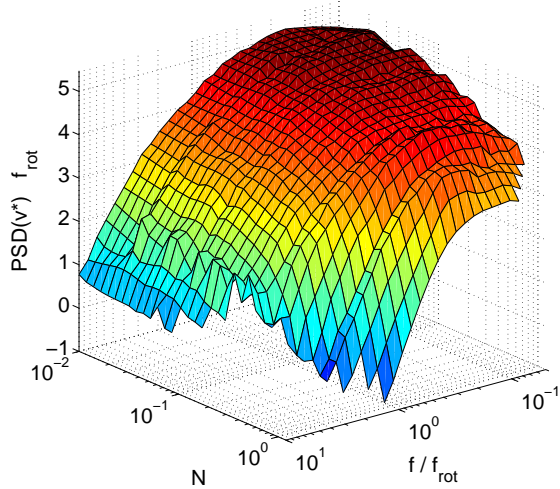}
\caption{\label{sp3D} Evolution of the shape of the power spectrum of the normalized potential as a function of the interaction parameter. Data for 5, 10, 15 and 20~Hz have been used for this representation.}
\end{figure}
Using the full ensemble of datasets at the various rotation rates one can interpolate the evolution of the shape of the dimensionless spectra as a function of both $f/f_{rot}$ and $N$. The result is shown in figure~\ref{sp3D}. This representation summarizes all previous observations. As $N$ increases, first there is a self similar decay of the spectra. When $N$ reaches approximately 0.1, an additional specific damping of the high frequencies is observed. The high frequency spectrum gets extremely steep. This extremely steep regime covers the full inertial range for $N$ close to 1.


\subsection{Induced magnetic field}

\begin{figure}[!htb]
(a)\includegraphics[width=7cm]{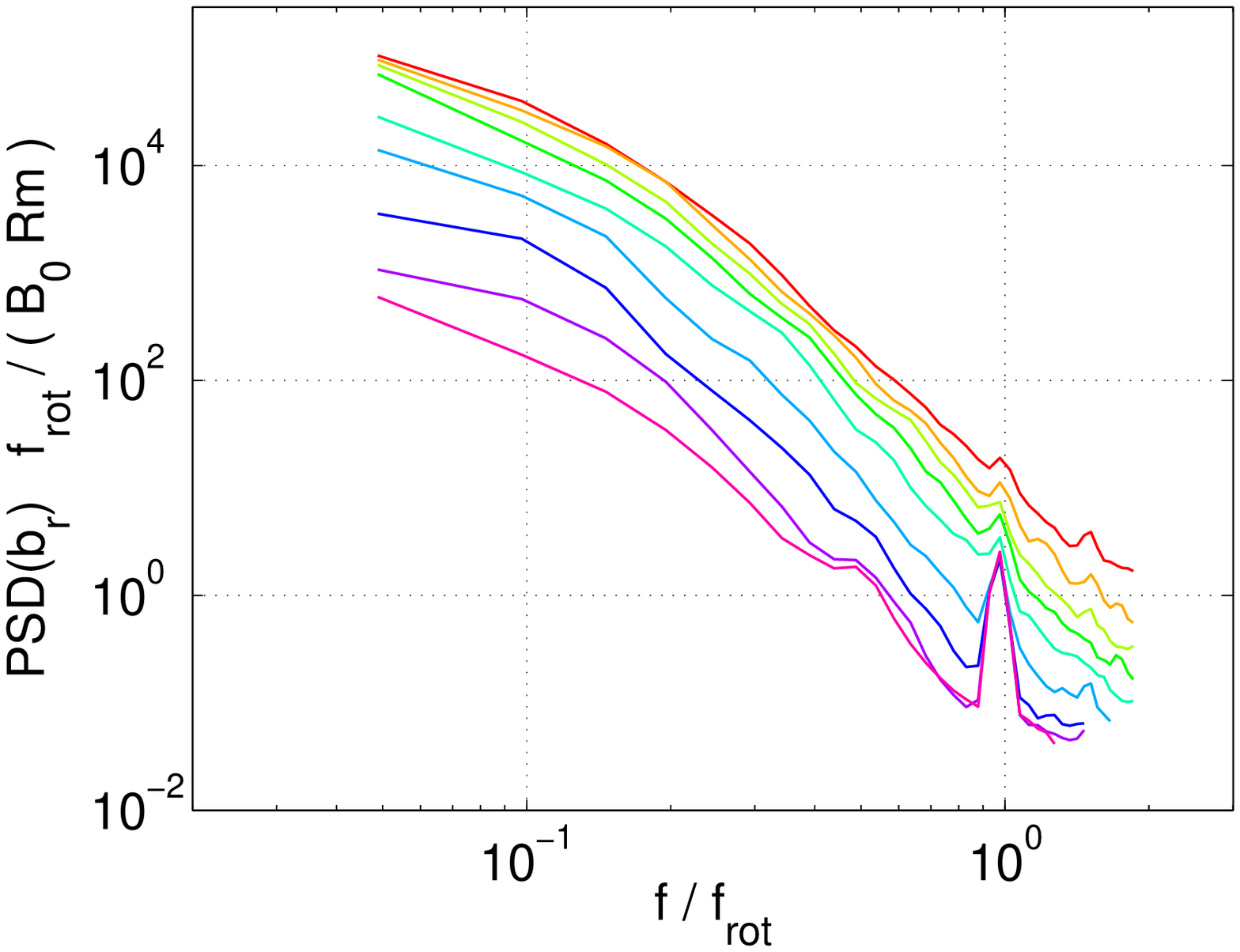}
(b)\includegraphics[width=7cm]{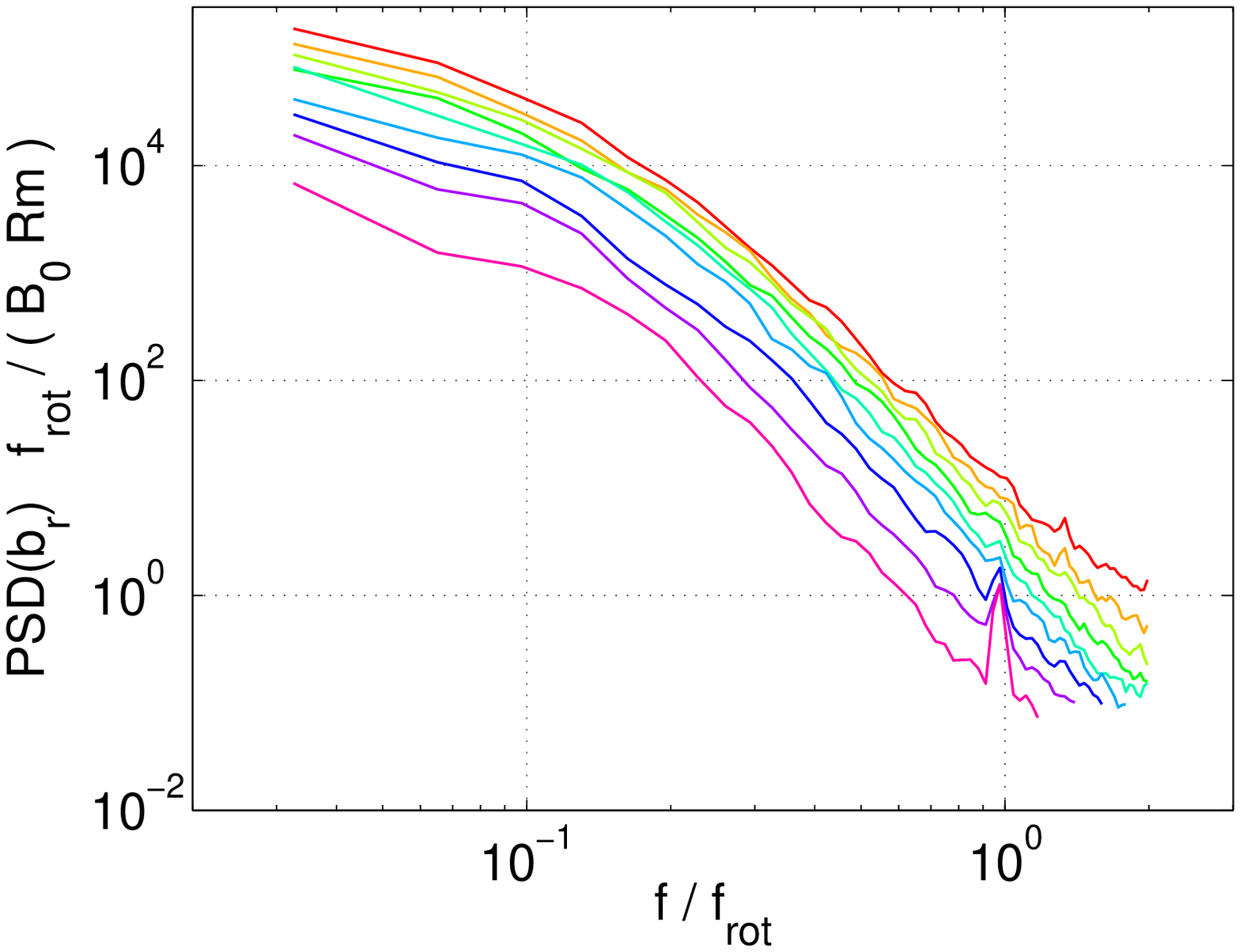}
(c)\includegraphics[width=7cm]{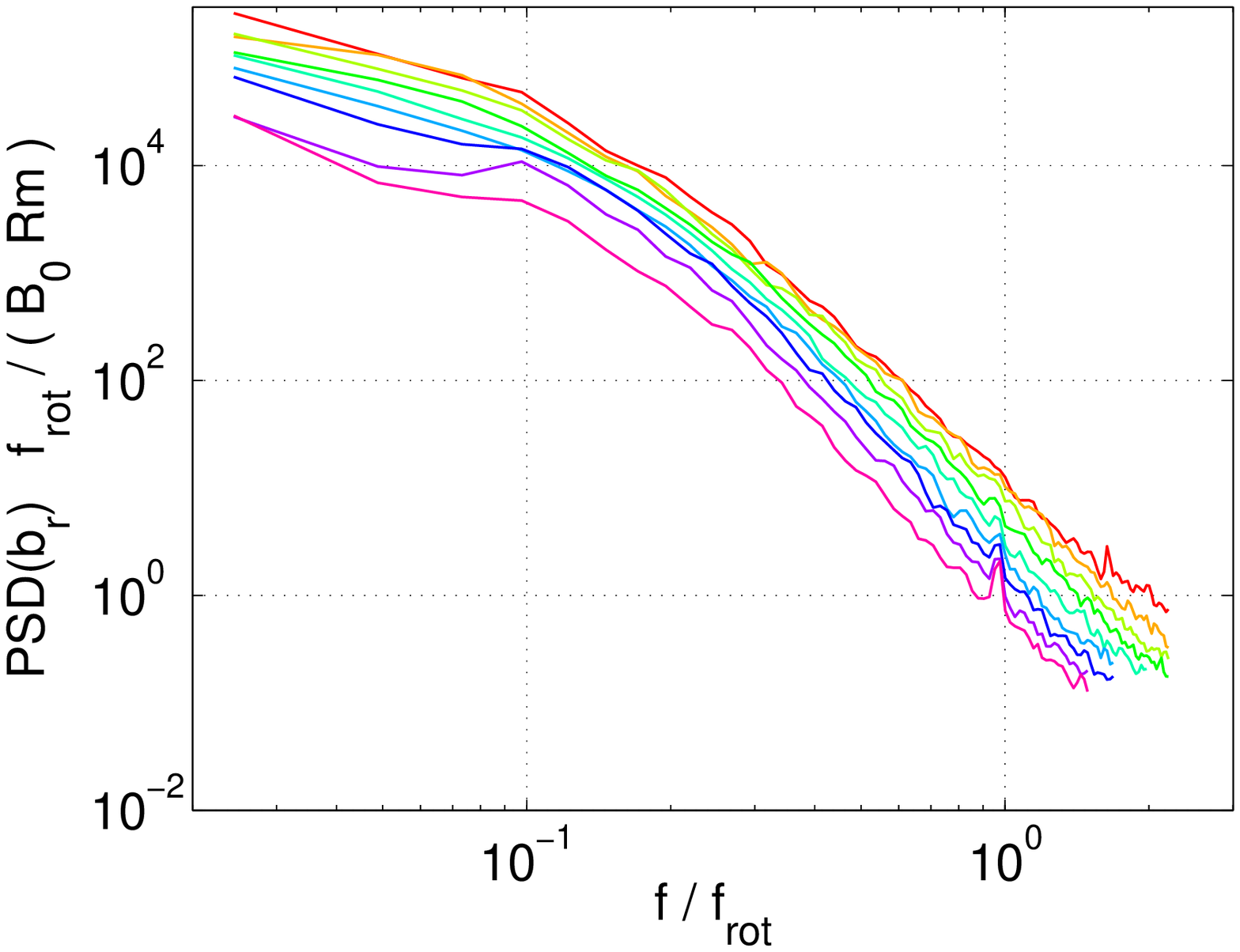}
\caption{\label{spb} Evolution of the power spectrum of the dimensionless induced magnetic field $\dfrac{b_r}{B_0 Rm}$. The subfigures correspond to different rotation rates of the propellers: 
(a) 10~Hz, (b) 15~Hz and (c) 20~Hz. 
In each subfigure, the various curves correspond to different applied magnetic fields. The curves are naturally ordered from top to bottom as the magnetic field is increased. The results for the following values of $B_0$ are displayed: $B_0=178$, 356, 534, 712, 890, 1070, 1250, 1420 and 1600~G. The noise part of the spectra has been removed for the clarity of the figures.}
\end{figure}
The same analysis can be performed on the induced magnetic field. The dimensionless spectra of $b_r$ are shown in figure~\ref{spb}. One concern is that the induced magnetic field is low, so that the signal to noise ratio of the gaussmeter is not as good as that of the potential probes. The magnetic field spectrum reaches the noise level at a frequency which is approximately $2f_{rot}$. This is also related to the fact that the scaling law expected from a Kolmogorov-like analysis for the magnetic field is much steeper than the one of the velocity ($f^{-11/3}$ in the dissipative range of the magnetic field, at frequencies corresponding to the inertial range of a velocity scaling as $f^{-5/3}$). 
Nevertheless, the same two regimes are observed: at low $N$ the spectra are damped in a self similar way. At large $N$ the high frequencies seem to be overdamped. However, the picture is not so clear because of the low signal to noise ratio.

\begin{figure}[!htb]
(a)\includegraphics[width=7cm]{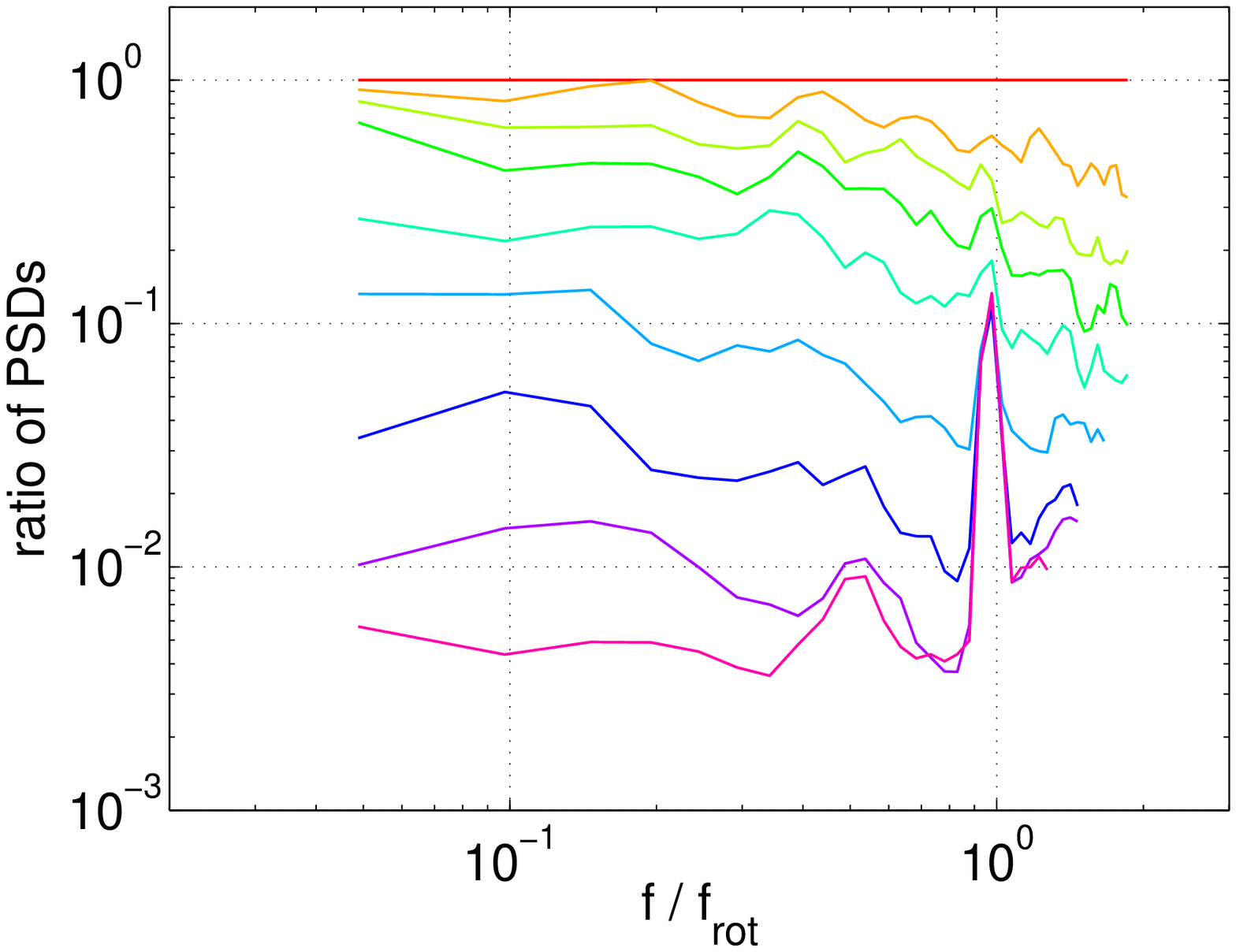}
(b)\includegraphics[width=7cm]{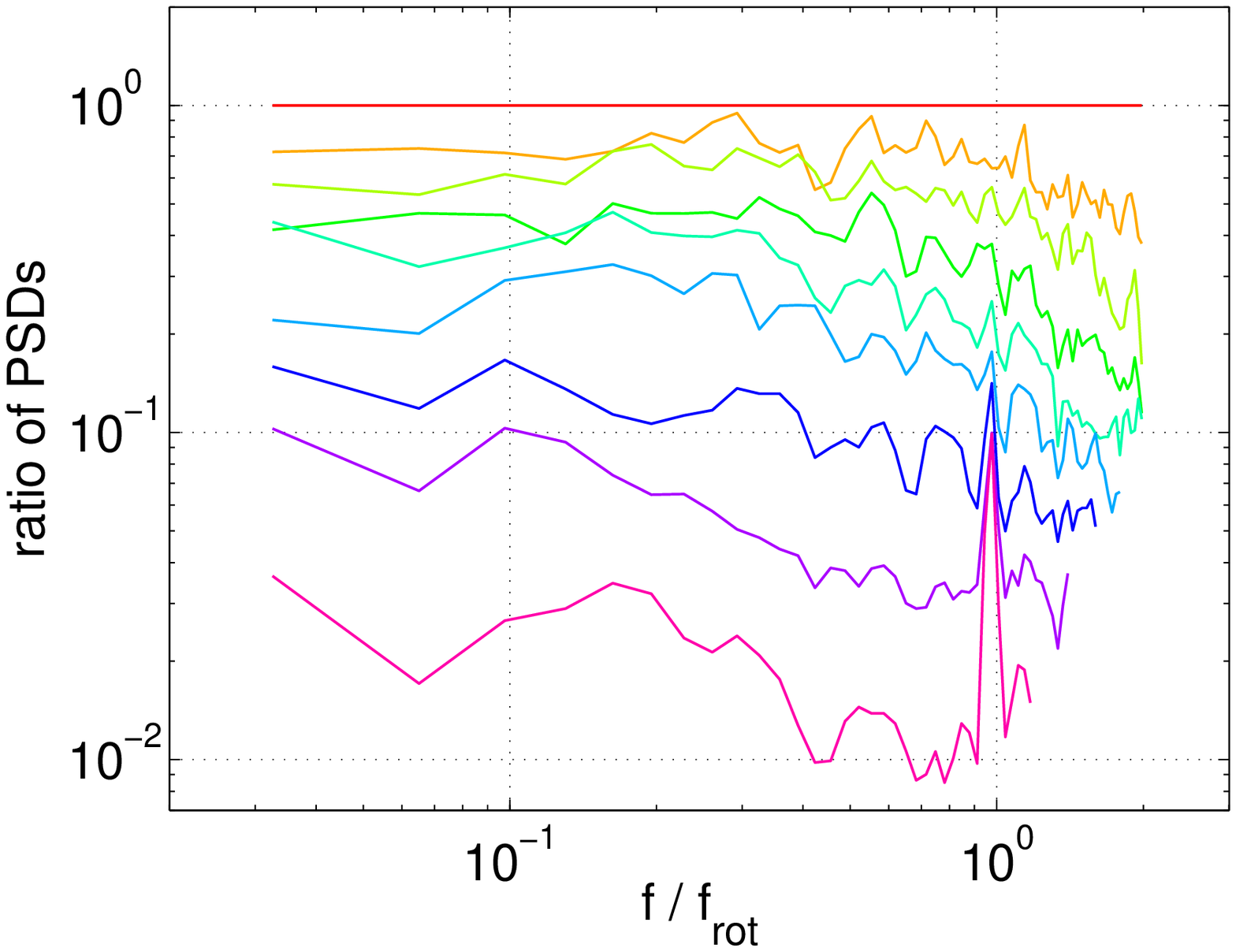}
(c)\includegraphics[width=7cm]{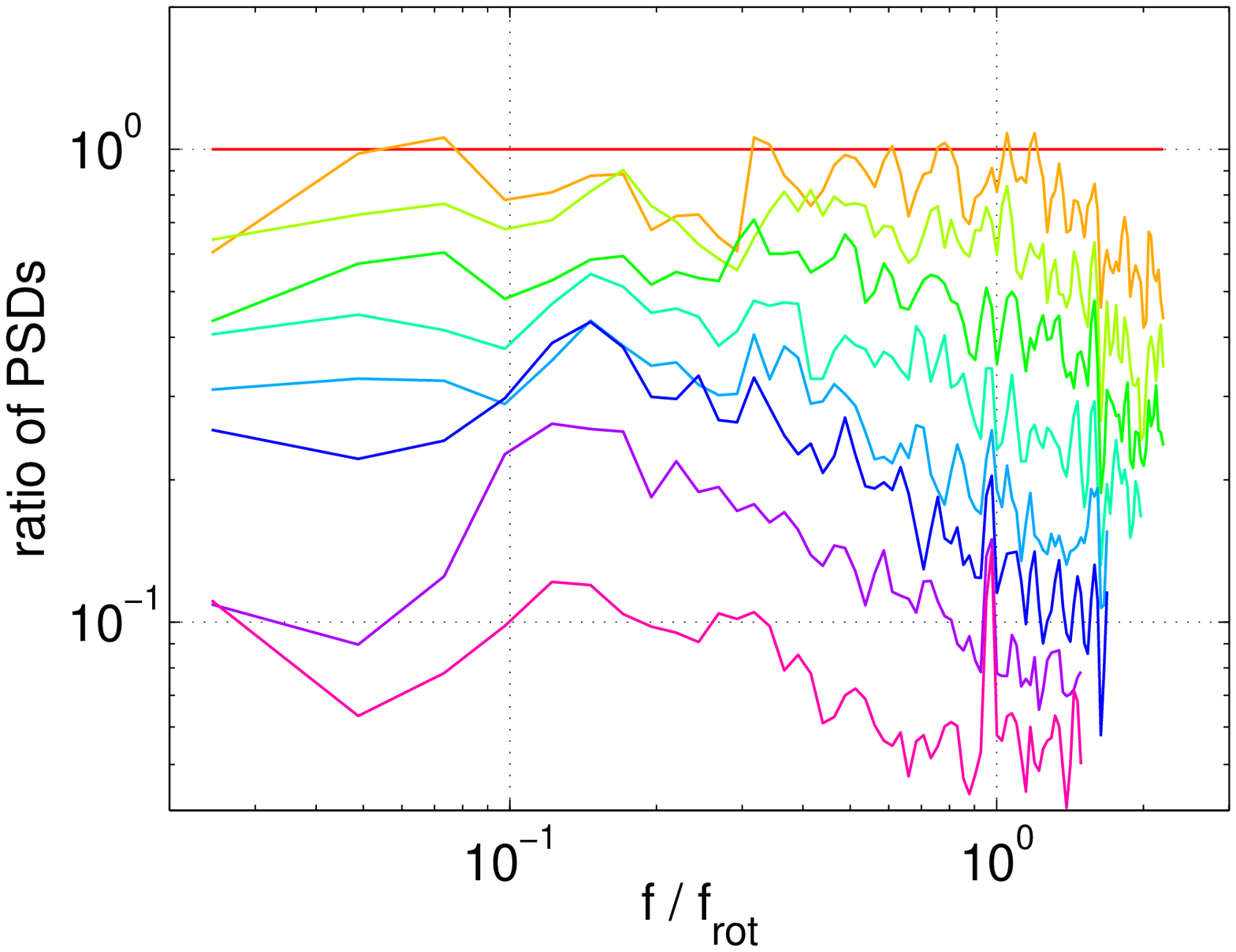}
\caption{\label{spbn} Evolution of the shape of the power spectrum of the induced magnetic field $b_r$. The spectra of figure~\ref{spb} are divided by the spectrum obtained at the smallest value of the applied magnetic field $B_0=178$~G and at the same $f_{rot}$. The subfigures correspond to different rotation rates of the propellers: 
(a) 10~Hz, (b) 15~Hz and (c) 20~Hz.}
\end{figure}
This last point is better seen on the ratio of the spectra in figure~\ref{spbn}. The ratio corresponding to the highest applied magnetic field is decaying at large frequencies.

\subsection{Coherence between velocity and magnetic field}
\begin{figure}[!htb]
(a)\includegraphics[width=7cm]{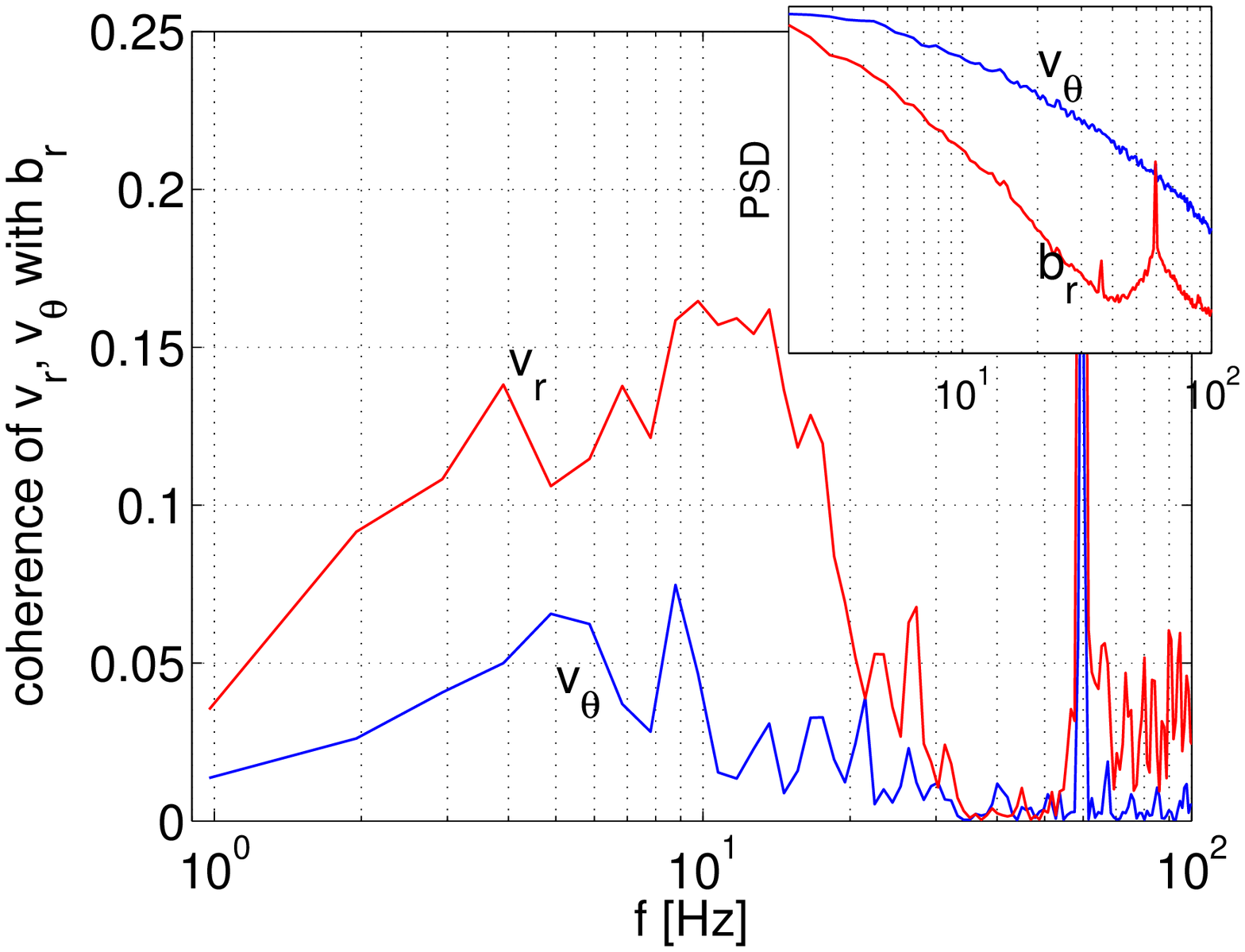}
(b)\includegraphics[width=7cm]{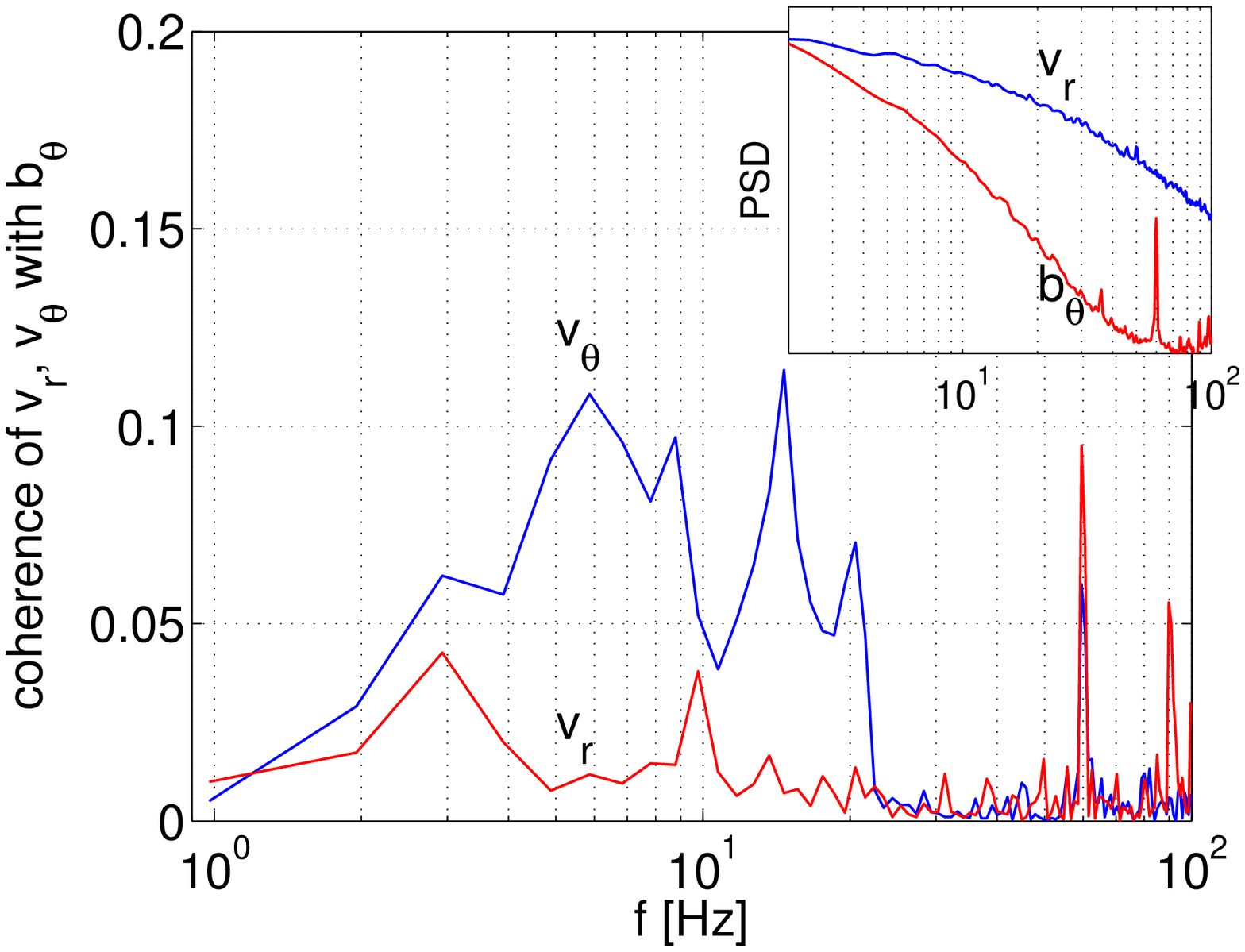}
\caption{\label{coh} Spectral coherence between the velocity and the induced magnetic field. (a) coherence between $b_r$ and $v_r$ or $v_\theta$. The inset recalls the spectra of $b_r$ and $v_\theta$. (b) coherence between $b_\theta$ and $v_r$ or $v_\theta$. The inset recalls the spectra of $b_\theta$ and $v_r$. The data correspond to $f_{rot}=15$~Hz and $B_0=356$~G.}
\end{figure}
The experiment has been designed to record the potential difference and the magnetic field in the vicinity of the same point. Figure~\ref{coh} displays the spectral coherence between these two quantities. In this figure we have identified the azimuthal potential difference with the radial component of velocity $v_r$ and the radial potential difference with the azimuthal component of velocity $v_\theta$, even though this identification may be somewhat abusive. We recall that the coherence is one when both signals are fully correlated at a given frequency and zero if they are uncorrelated at this frequency. We have plotted in the insets the power spectrum of each signal. The magnetic field spectrum falls below the noise level at about $2f_{rot}$, which explains why all coherence curves go to zero above $2f_{rot}$ (there is no magnetic signal at these frequencies). For lower frequencies, a small but non-zero value of the coherence is observed for the following pairs: $(b_r,v_r)$ with a coherence level above 0.1, $(b_r,v_\theta)$ with a coherence close to 0.05 and $(b_\theta,v_\theta)$ with a coherence level barely reaching 0.1. The coherence value changes a little for other values of $f_{rot}$, but the global picture remains the same, with $(b_r,v_r)$ being the most coherent.

\begin{figure}[!htb]
\includegraphics[width=8cm]{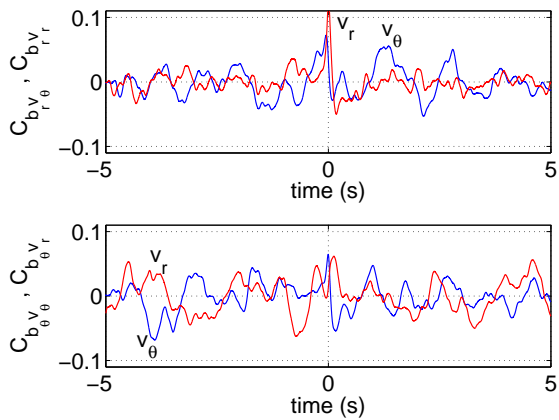}
\caption{\label{cor} Correlation coefficients between the velocity and the induced magnetic field. (a) Correlation between $b_r$ and $v_r$ (upper curve) or $v_\theta$ (lower curve). (b) Correlation between $b_\theta$ and $v_r$ (lower curve) or $v_r$ (upper curve). The data correspond to $f_{rot}=15$~Hz and $B_0=356$~G.}
\end{figure}
The time correlations are shown in figure~\ref{cor}. No clear correlation is observed except for the pair $(b_r,v_r)$ which displays a small peak reaching 0.1. The weak coherence observed in the previous figure is barely seen here, probably because of an insufficient convergence of the correlation function.

At the low values of magnetic Reynolds number attained in our experiment, the magnetic field is diffused through Joule effect, so that its structure is mainly large scale. The Reynolds number being large, the velocity fluctuations develop down to much smaller scales. Because of the strong ohmic diffusion, the magnetic field is expected to be sensitive mostly to large-scale and low frequency fluctuations of the velocity field. For example one could expect to observe a correlation or coherence between $v_\theta$ and $b_\theta$: fluctuations of the differential rotation can induce fluctuations in the conversion of the axial $\mathbf B_0$ into azimuthal magnetic field through $\omega$ effect. As far as $b_r$ is concerned, the poloidal recirculation bends the vertical field lines towards the exterior of the tank  in the vicinity of the mid-plane, a process which induces radial magnetic field from the vertical applied field $\mathbf B_0$. Fluctuations of this poloidal recirculation thus directly impacts the radial induced magnetic field, hence the correlation between $v_r$ and $b_r$. Although such low frequency coherence is indeed observed in figure \ref{coh} its amplitude remains very low, so that there is almost no correlation between the velocity field and the magnetic field measured in the vicinity of the same point.

\section{Discussion and conclusion}
The effect of a strong magnetic field on forced turbulence is studied experimentally with potential probes and induced magnetic field mesurements. The velocity fluctuations are strongly damped as the applied magnetic field increases: for $N \simeq 0.6$,  the turbulence intensity in the mid-plane of the tank is decreased by an order of magnitude. As a consequence, the standard deviation of the induced magnetic field - normalized by $B_0 Rm$ - also diminishes by a factor ten. The spectrum of the non-dimensional potential $v^*$ is affected in two different ways by the magnetic field: for low values of the interaction parameter the spectrum decays uniformly at all frequencies, its shape remaining the same. For higher values of $N$, we observe an overdamping of the high frequencies. The same effect is seen on the induced magnetic field spectra.

We identify several features which highlight the very different behaviors of forced and decaying turbulence when they are subject to a strong magnetic field: decaying turbulence is thought to evolve towards a bidimensional structure. Its velocity spectrum displays a $-3$ exponent which can be attributed either to this bidimensionalization or to a quasi-steady equilibrium between velocity transfer and ohmic dissipation. In the present experiment, no such $-3$ exponent is observed, and the spectra are much steeper (exponent $-5$ to $-6$) for values of $N$ higher than $0.5$.
Moreover, we have introduced a parameter $a=\eta \frac{b_{rms}}{\delta\phi_{rms}}$ to quantify the anisotropy of the turbulence in the mid-plane of the tank. When $N$ increases, the decrease of this parameter by a factor 3 is the signature of the elongation of the flow structures along the applied magnetic field. However, the flow always remains 3D since $a$ is non-zero even for the highest value of the interaction parameter reached in this experiment. These differences between forced and decaying turbulence come from the 3D forcing imposed by the propellers, which rules out the possibility of a 2D statistically steady state of the flow. It would be interesting to perform the same kind of experimental study of the anisotropy with other forcing mechanisms, such as current-driven MHD flows or turbulent thermal convection in a liquid metal. 

We have studied the evolution of the injected mechanical power as the applied magnetic field increases and found almost no influence of the latter: the injected mechanical power remains the same although velocity fluctuations are decreased by an order of magnitude in the central shear layer.

Finally, we stress the poor level of correlation between the velocity and induced magnetic fields measured in the vicinity of the same point, and attribute it to the scale separation between the two fields.

The strong damping of turbulent fluctuations by the magnetic field can be invoked as a saturation mechanism for turbulent dynamos: in a dynamo experiment, one observes spontaneous generation of magnetic field when the magnetic Reynolds number is above a critical value $Rm_c$. If the turbulent fluctuations are involved in the generating process of the magnetic field (through $\alpha\omega$ or $\alpha^2$ mechanisms for instance), there is a critical level of rms turbulent fluctuations $\sigma_{vc}$ above which magnetic field is generated ($\sigma_v=\sigma_{vc}$ for $Rm=Rm_c$). For $Rm > Rm_c$, the initial level of turbulent fluctuations is above $\sigma_{vc}$, and the magnetic field grows exponentially from a small perturbation: the interaction parameter increases, and the turbulent fluctuations are damped according to figure \ref{sigvstar}. An equilibrium is reached when  the damping is such that the rms turbulent fluctuations are reduced to $\sigma_{vc}$. For small values of $N$ we observed that $\sigma_{v} (N)=\sigma_{v}(N=0) e^{-\gamma N} \simeq \sigma_{v}(N=0) (1-\gamma N) $, with $2.5<\gamma<3.5$. The saturated value $N_{sat}$ of $N$ follows from the equality $\sigma_{v} (N_{sat}) = \sigma_{vc} = \sigma_{v}(N=0) (1-\gamma N_{sat})$. As $\sigma_{v}$ is proportional to $Rm$, we get $N_{sat}=\frac{Rm-Rm_c}{\gamma Rm_c}$, hence the following scaling law for the magnetic field:
\begin{equation}
B_{sat}^2 =\frac{2 \pi \rho R f_{rot}}{\sigma L \gamma} \frac{Rm-Rm_c}{Rm_c}
\end{equation}
This is the turbulent scaling law for the saturation of a dynamo, which was originally described by P\'etr\'elis et al. (see \cite{Petrelis} for instance). Although the exact geometry of the experimental setup and large scale magnetic fields are different from that of the present experimental study, results from the VKS dynamo can be used to test this relationship: using $R=15.5$~cm, $L=21$~cm, $f_{rot}=16$~Hz, $\rho=930$~kgm$^{-3}$, $\sigma=9.5\,10^6$~$\Omega^{-1} m^{-1}$ and $\gamma=3$, the computed magnetic field amplitude is $B_{sat} \simeq 290$~G for $\frac{Rm-Rm_c}{Rm_c}=\frac{1}{3}$. This is the right order of magnitude: the amplitude of the VKS dynamo field measured in the vicinity of the axis of the cylinder is approximately $150 G$ for this value of $Rm$ (\cite{VKSP5}: figure 3(b)).

The authors would like to thank F. P\'etr\'elis for his comments and for his help in the design of the experimental setup, and S Fauve for insightful discussions. This work is supported by
ANR BLAN08-2-337433.


\begin{thebibliography}{1}

\bibitem{Eckert}
{S. Eckert, G. Gerbeth, W. Witke and H. Langenbrunner},
\newblock {``MHD turbulence measurements in a sodium channel flow exposed to a transverse magnetic field,''}
\newblock Int. J. Heat Fluid Flow \textbf{22}, p. 358-364, (2001).

\bibitem{Alemany}
{A. Alemany, R. Moreau, P.L. Sulem and U. Frisch},
\newblock {``Influence of an external magnetic field on homogeneous MHD turbulence,''}
\newblock J. M\'ecanique \textbf{18}, 2, (1979).

\bibitem{Knaepen}
{B. Knaepen, R. Moreau},
\newblock {\em Magnetohydrodynamic turbulence at low magnetic Reynolds number},
\newblock Ann. Rev. Fluid Mech. \textbf{40}, p. 25-45 (2008).

\bibitem{Sisan}
{D.~R. Sisan, W.~L. Shew and D.~P. Lathrop},
\newblock {``Lorentz force effects in magneto-turbulence,''}
\newblock Phys. Earth Planet. Int.  \textbf{135}, p. 137-159, (2003).

\bibitem{Zikanov}
{O. Zikanov and A. Thess},
\newblock {``Direct numerical simulation of forced MHD turbulence at low magnetic Reynolds number,''}
\newblock J. Fluid Mech.  \textbf{358}, p. 299-333, (1998).

\bibitem{Zikanov2}
{T. Boeck, D. Krasnov, A. Thess and O. Zikanov},
\newblock {``Large-Scale Intermittency of Liquid-Metal Channel Flow in a Magnetic Field,''}
\newblock Phys. Rev. Lett.  \textbf{101}, 244501, (2008).

\bibitem{Vorobev}
{A. Vorobev, O. Zikanov, P.~A. Davidson and B. Knaepen},
\newblock {``Anisotropy of magnetohydrodynamic turbulence at low magnetic Reynolds number,''}
\newblock Phys. Fluids \textbf{17}, 125105, (2005).

\bibitem{Burattini}
{P. Burattini, M. Kinet, D. Carati and B. Knaepen},
\newblock {``Anisotropy of velocity spectra in quasistatic magnetohydrodynamic turbulence,''}
\newblock Phys. Fluids \textbf{20}, 065110, (2008).

\bibitem{Marie}
{L. Mari\'e, F. Daviaud},
\newblock {``Experimental measurement of the scale-by-scale momentum transport budget in a turbulent shear flow,''}
\newblock Phys. Fluids \textbf{16}, p. 457-461, (2004).

\bibitem{Monchaux}
{R. Monchaux, M. Berhanu, M. Bourgoin, M. Moulin,P. Odier, J.-F. Pinton, R. Volk, S. Fauve, N. Mordant, F. P\'etr\'elis, A. Chiffaudel, F. Daviaud, B. Dubrulle, C. Gasquet, L. Mari\'e and F. Ravelet},
\newblock {``Generation of a Magnetic Field by Dynamo Action in a Turbulent Flow of Liquid Sodium,''}
\newblock Phys. Rev. Lett.  \textbf{98}, 044502, (2007).

\bibitem{Moffatt}
{H.~K. Moffatt},
\newblock {\em Magnetic Field Generation in Electrically conducting Fluids},
\newblock (Cambridge University Press, 1978).

\bibitem{Kharicha}
{A. Kharicha, A. Alemany, D. Bornas},
\newblock {``Influence of the magnetic field and the conductance ratio on the mass transfer rotating lid driven flow,''}
\newblock Int. J. Heat Mass Transfer \textbf{47}, p. 1997-2014, (2004).

\bibitem{Ricou}
{R. Ricou and C. Vives},
\newblock {``Local velocity and mass transfer measurements in molten metals using an incorporated magnet probe,''}
\newblock Int. J. Heat Mass Transfer \textbf{25}, p. 1579-1588, (1982).

\bibitem{Tsinober}
{A. Tsinober, E. Kit, M. Teitel},
\newblock {``On the relevance of the potential-difference method for turbulence measurements,''}
\newblock J. Fluid Mech. \textbf{175}, p. 447-461, (1987).

\bibitem{Tennekes}
{H. Tennekes and J.L. Lumley},
\newblock {\em A First Course in Turbulence},
\newblock (MIT Press, Cambridge, 1972).

\bibitem{Bolonov}
{N.~I. Bolonov, A.~M. Kharenko, A.~E. \' Eidel'man},
\newblock {``Correction of spectrum of turbulence in the measurement by a conduction anemometer,''}
\newblock Inzhernerno-Fizicheskii Zhurnal  \textbf{31}, 243, (1976).

\bibitem{Berhanu}
{M. Berhanu, B. Gallet, N. Mordant and S. Fauve},
\newblock {``Reduction of velocity fluctuations in a turbulent flow of liquid gallium by an external magnetic field,''}
\newblock Phys. Rev. E \textbf{78}, 015302, (2008).

\bibitem{Odier}
{P. Odier, J.-F. Pinton and S. Fauve},
\newblock {``Advection of a magnetic field by a turbulent swirling flow,''}
\newblock Phys. Rev. E \textbf{58}, p. 7397-7401, (1998).

\bibitem{Sommeria}
{J. Sommeria and R. Moreau},
\newblock {``Why, how, and when, MHD turbulence becomes two-dimensional,''}
\newblock J. Fluid Mech. \textbf{118}, p. 507-518, (1982).

\bibitem{Petrelis}
{F. P\'etr\'elis, N. Mordant, S. Fauve},
\newblock {``On the magnetic fields generated by experimental dynamos,''}
\newblock Geophysical and Astrophysical Fluid Dynamics  \textbf{101}, p. 289-323, (2007).

\bibitem{VKSP5}
{R. Monchaux, M. Berhanu, S. Auma\^itre, A. Chiffaudel, F. Daviaud, B. Dubrulle, F. Ravelet, S. Fauve, N. Mordant, F. P\'etr\'elis,
M. Bourgoin, P. Odier, J.-F. Pinton, N. Plihon and R. Volk},
\newblock {``The Von K\'arm\'an Sodium experiment: turbulent dynamical dynamos,''}
\newblock Phys. Fluids, \textbf{21}, 035108 , (2009).

\end{thebibliography}
\end{document}